\documentclass[reprint,
 superscriptaddress,
 numerical,
 showpacs,
 amsmath,amssymb,
 aps,
 longbibliography,
 prb, floatfix
]{revtex4-2}
\usepackage{graphicx}

\usepackage[colorlinks=true,
  urlcolor=blue,
  linkcolor=blue,
  citecolor=blue
]{hyperref}

\usepackage[pdftex]{color}
\usepackage[usenames,dvipsnames,svgnames,table]{xcolor}

\definecolor{Red_Nice}{HTML}{DC322F}
\definecolor{Blue_Nice}{HTML}{268BD2}
\definecolor{Green_Nice}{HTML}{859900}

\usepackage{mathdots}
\usepackage{mathtools}

\usepackage{txfonts}
\usepackage{newtxmath}
\usepackage{newtxtext}

\usepackage{textcomp}

\usepackage{multirow}

\newcommand{\ket}[1]{| #1 \rangle}
\newcommand{\bra}[1]{\langle#1 |}

\DeclareMathOperator{\diag}{diag}
\DeclareMathOperator{\Pf}{Pf}
\DeclareMathOperator{\Tr}{Tr}
\DeclareMathOperator{\arccosh}{arccosh}
\DeclareMathOperator{\sech}{sech}

\begin{document}

\title{Quantum tunneling in the presence of a topology-changing fermionic bath}

\author{Elis Roberts}
\affiliation{T.C.M. Group, Cavendish Laboratory, University of Cambridge, J.J. Thomson Avenue, Cambridge, CB3 0HE, United Kingdom}
\author{Jan Behrends}
\affiliation{T.C.M. Group, Cavendish Laboratory, University of Cambridge, J.J. Thomson Avenue, Cambridge, CB3 0HE, United Kingdom}
\author{Benjamin B\'{e}ri}
\affiliation{T.C.M. Group, Cavendish Laboratory, University of Cambridge, J.J. Thomson Avenue, Cambridge, CB3 0HE, United Kingdom}
\affiliation{DAMTP, University of Cambridge, Wilberforce Road, Cambridge, CB3 0WA, United Kingdom}

\begin{abstract}
Coupling a quantum particle to a fermionic bath suppresses the particle's amplitude to tunnel, even at zero temperature. 
While this effect can generally be neglected for gapped baths---a key feature for superconducting qubits---, it is possible for the bath to be gapped near the potential minima between which the particle tunnels, but different minima to correspond to different bath topologies.  
This enforces the bath to undergo gap closing along the tunneling path. 
In this work, we investigate quantum tunneling in the presence of such a topology-changing fermionic bath.
We develop a field theory for this problem, linking the instantons describing tunneling in a bath of $d$ space dimensions to topological boundary modes of systems in $d+1$ dimensions, thus stepping a level higher in a dimensional hierarchy.
We study in detail a $d=1$ example, inspired by planar Josephson junctions where the particle coordinate is the superconducting phase whose value sets the electronic topology. 
We find that the topology change suppresses tunneling by a factor scaling exponentially with the system size. 
This translates to a correspondingly enhanced suppression of the energy splitting for the lowest-lying states, despite these being linear combinations of states near potential minima where the bath is gapped.
Our results help to estimate the influence of charging energy on topological phases arising due to the Josephson effect and, conversely, to assess the potential utility of such topological systems as superconducting qubits.
For moderate-sized baths, the incomplete suppression of tunneling opens the prospects of quantum-mechanical superpositions of many-body states of different topology, including superpositions of states with and without Majorana fermions. 
\end{abstract}

\maketitle

\section{Introduction}

Quantum tunneling of a particle can be significantly altered in the presence of a bath.
At zero temperature, a gapless bath reduces the tunneling amplitude by a factor that is exponentially small in the system-to-bath coupling and the tunneling potential width~\cite{Caldeira:1981bf,Caldeira:1983jq}.
For gapped baths, however, the exponent decreases with the ratio of the bath's gap and the oscillation frequency characterizing the minima of the tunneling potential~\cite{Grabert:1984gk,Dorsey:1985iw,Legget:1987di}, and so the bath can be neglected for large gaps.
This feature is crucial for the coherence of Josephson-junction-based superconducting qubits~\cite{Ambegaokar:1982do,Larkin:1983cy,Legget:1987di,Makhlin:2001ba,Koch:2007gz}, where the position of the tunneling ``particle'' is the superconducting phase difference $\phi$; in these systems the gapped fermionic bath of electrons merely renormalizes the junction's capacitance~\cite{Kampf:1988jw,Schon:1990kj}.

In-gap fermionic levels change this picture~\cite{Kampf:1988jw,Badiane:2013gc}:
For example, an approximate~\cite{Averin_PhysRevLett.82.3685,Bargerbos_PhysRevLett.124.246802} or symmetry-enforced~\cite{Kitaev:2001gb,kwon2004fractional,Fu_PhysRevB.79.161408,Heck_PhysRevB.84.180502,pekker_suppression_2013,Badiane:2013gc} crossing of in-gap levels along the tunneling path acts to suppress quantum tunneling.
In this work, we consider a more dramatic scenario: 
What happens if the bath is gapped at the minima of the tunneling potential, but 
undergoes a \emph{bulk gap closing} (i.e., 
merging of level continua, instead of a single level crossing) when the particle tunnels between the minima?

Such gap closings can be enforced by topology:
When the bath Hamiltonian depends on the particle coordinate such that potential minima correspond to gapped bath Hamiltonians but with different minima corresponding to different values of a suitable Hamiltonian topological invariant~\cite{HasanKane_RevModPhys.82.3045,QiZhang_RevModPhys.83.1057,bernevig2013topological,asboth2016short}, then the robustness of this invariant under deformations that do not close the bulk gap
implies that the bath must undergo a bulk gap closing somewhere along the tunneling path. 
This scenario is different from previous works on gauge theories with topologically distinct vacua~\cite{Belavin:1975cz,Callan:1976ef,coleman_aspects_1985,rajaraman_instanton_book}: there, topology is that of gauge field configurations, i.e., of instantons, while for us the crucial form topology is fermionic, in the sense of topological insulators and superconductors~\cite{HasanKane_RevModPhys.82.3045,QiZhang_RevModPhys.83.1057,bernevig2013topological}.

Instantons, however, also enter the scenario we aim to study, as they provide a general field theoretical framework for tunneling amplitudes~\cite{coleman_aspects_1985,rajaraman_instanton_book}.
In this field theoretical language, the problem we are interested in corresponds to, as we shall explain, instantons linking the topology change of a $d$-dimensional bath to protected gapless modes in interfaces between topologically distinct phases in $d+1$ dimensions. 

The question we set to answer is inspired by theoretical~\cite{pientka_topological_2017,Hell:2017fn} and experimental~\cite{Fornieri:2019hz,Ren:2019ew} work on planar Josephson junctions, motivating the $d=1$ example that we shall be mostly focusing on.
In these setups, the fermionic topology changes as a function of the superconducting phase difference $\phi$ across the junction~\cite{pientka_topological_2017,Hell:2017fn}, resulting in two topologically inequivalent minima of the effective Josephson potential.
The concept we investigate is however more general and can arise in other settings where the control parameter for whether a fermionic system is in a topological phase can be promoted into a quantum variable.

In the language of planar Josephson junctions, the new ingredient we add is charging energy, a contribution to the Hamiltonian to be considered, e.g., in Cooper pair box systems~\cite{nazarov2009quantum}.
Charging energy serves as a kinetic term for $\phi$, hence it is its presence that enables $\phi$ to quantum tunnel between the minima of the effective Josephson potential it experiences. 

Our main finding is that the topology-enforced gap closing reduces the tunneling amplitude exponentially in the size of the bath (which for our planar-junction inspired model is also the system size).
We derive, and quantify, this result using instantons, a method which we employ both analytically in illuminating limits, and in variational numerics in more general cases.
We find that our variational approach, although focused on a certain instanton ansatz, performs excellently, as confirmed by comparisons with the ground state energy splitting obtained by exact diagonalization.

The rest of the paper is organized as follows:
After introducing our $d=1$ model in Sec.~\ref{sec:model}, we describe the field theory for the general problem in Sec.~\ref{sec:instantons}. 
We then link instantons coupled to $d$-dimensional fermions to topological boundary modes of systems in $d+1$ dimension in Sec.~\ref{subsec:low_energy_analogy}, a section which also includes a detailed analytical study of our $d=1$ model, focusing on the  ``sharp-instanton limit'' to illuminate key topological features.
We describe our variational approach in Sec.~\ref{sec:variational} and compare our instanton-based results with exact diagonalization in Sec.~\ref{sec:results}. 
We summarize our results and discuss some implications and generalizations in Sec.~\ref{sec:conclusions}.

\section{Model}
\label{sec:model}

Our model is inspired by proposals that implement a nontrivial topological superconductor in planar Josephson junctions~\cite{Hell:2017fn,pientka_topological_2017}.
In these quasi-one-dimensional systems at the interface between two superconductors, the phase difference $\phi$ between the superconductors drives a transition between topologically trivial and nontrivial regimes, with the latter hosting zero-energy Majorana end modes.
Importantly, the ground state energy forms a potential landscape with two topologically inequivalent minima~\cite{pientka_topological_2017}.
This behavior is captured by the effective Hamiltonian
\begin{equation}
 \mathcal{H}_{\phi}(k) = \Delta \cos^2 \left( \frac{\phi}{2} \right) \sigma_3 - \Delta \sin^2 \left( \frac{\phi}{2} \right) \left[ \cos(k) \sigma_3 + \sin(k) \sigma_2 \right],
 \label{eq:fermionic_hamiltonian}
\end{equation}
which is topologically trivial around $\phi =0$ and nontrivial~\cite{Kitaev:2001gb} around $\phi = \pi$, with transitions occurring at $\phi = \pi/2,3\pi/2$.
It interpolates between the two dimerized limits of the Kitaev chain at $\phi=0$ and $\phi=\pi$, where Majoranas are coupled either only on the same site or only between neighboring sites, respectively~\cite{Kitaev:2001gb}.
The Pauli matrices $\sigma_\mu$ act in particle-hole (PH) space and $\mathcal{H}_\phi$ respects both PH (with $\Xi = \sigma_1 \mathcal{K}$, where $\mathcal{K}$ is complex conjugation) and time-reversal symmetry (with $T= \mathcal{K}$).
The first term proportional to $\cos^2 (\phi/2)$ is a chemical potential, whose value we chose to match the superconducting order parameter $\Delta$.
The single-particle energies $\varepsilon_{\phi}^{\pm} (k) = \pm \Delta \sqrt{1 - \sin^2{(\phi)} \cos^2 (k/2)}$ equal $\pm \Delta$ at both $\phi = 0 $ and $\phi = \pi$.   

\begin{figure}
\includegraphics[width=\linewidth]{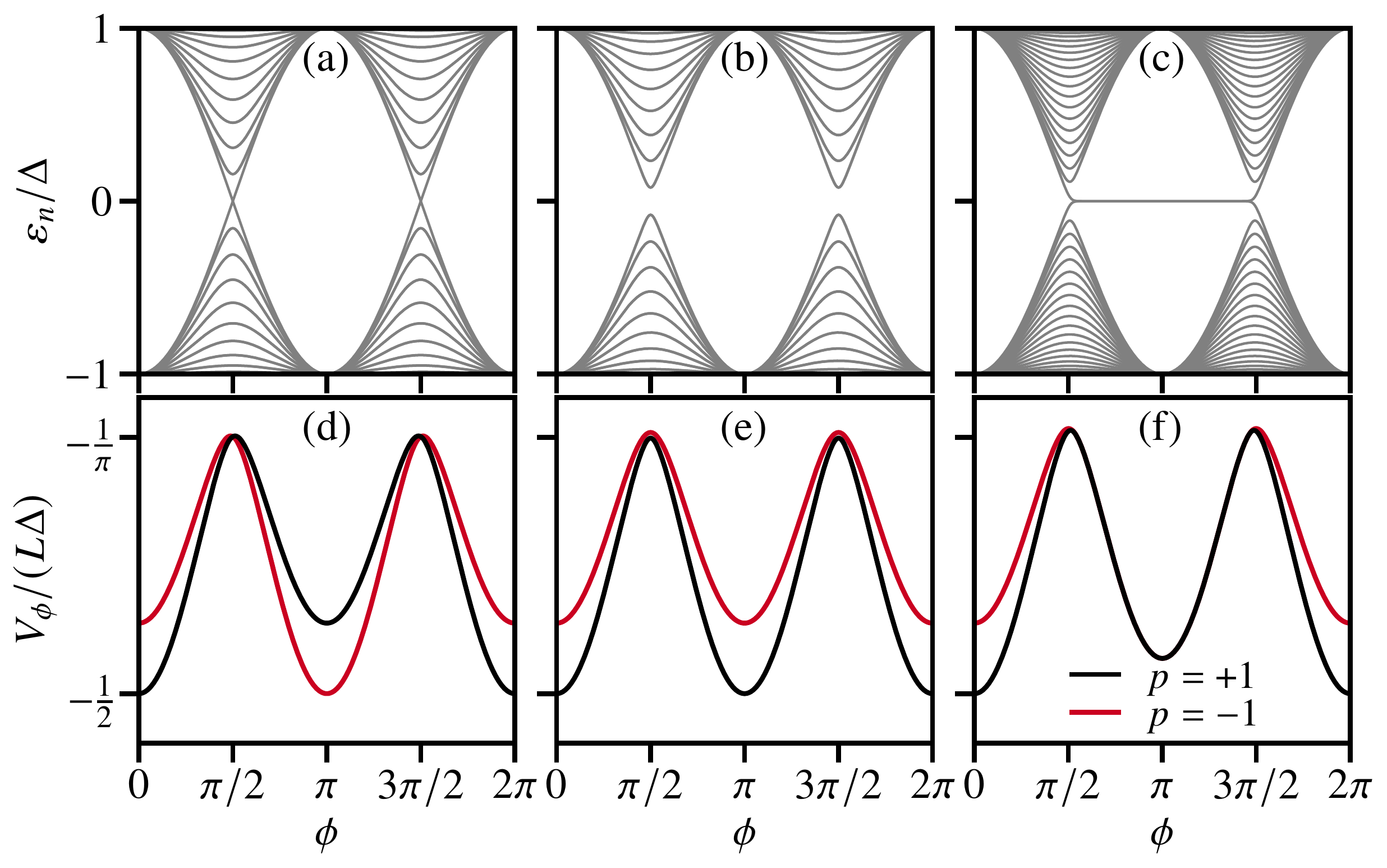}
\caption{(a)--(c) Single-particle energies of the fermionic Hamiltonian \eqref{eq:fermionic_hamiltonian} for $L=20$ sites.
(a) For periodic boundary conditions, the single-particle gap closes, while (b) for anti-periodic boundary conditions, a gap of order $1/L$ remains; (c) for open boundaries,  Majorana zero mode end states exist in the topological regime for $\pi/2 < \phi <3\pi/2$; their energy splitting vanishes exactly for $\phi=\pi$.
(d)--(f) Energies of the ground state and the first excited state, with the color denoting even (black, $p=+1$) and odd (red, $p=-1$) fermion parity.
(d) For periodic boundary conditions, the ground state parity is different in trivial and nontrivial sectors, while (e) the ground state parity is the same for anti-periodic boundaries; (f) for open boundaries, even and odd parities are nearly degenerate in the topological regime and exactly degenerate at $\phi = \pi$.
}
\label{fig:energies}
\end{figure}

We show the single-particle and ground state energies $V_\phi = \frac{1}{2} \sum_k \varepsilon_{\phi}^{-}(k)$ for different boundary conditions in Fig.~\ref{fig:energies}.
For periodic boundary conditions (PBC), the ground state parity changes a function of $\phi$, whereas for antiperiodic boundary conditions (APBC) the ground state remains in the same parity sector.
With open boundary conditions (OBC), Majorana zero modes form at the ends of the chain in the topological phase such that the ground state becomes approximately degenerate (with exponentially small splitting that disappears at $\phi=\pi$)~\cite{Kitaev:2001gb}.
While OBC are conceptually closest to existing experimental planar Josephson junction setups~\cite{Fornieri:2019hz,Ren:2019ew}, our model~\eqref{eq:fermionic_hamiltonian} can serve as a prototypical example for tunneling between topologically distinct phases also for the other boundary conditions. 

A key ingredient to our considerations is a kinetic term for $\phi$, motivated by the  charging energy in superconducting islands~\cite{Makhlin:2001ba}.
Including this, the full Hamiltonian governing the fermionic $c_k$ modes and the bosonic phase mode $\phi$ reads
\begin{equation}
 H = E_{\text{C}} (N - N_g)^2 + \frac{1}{2} \sum_k \Psi_k^\dagger \mathcal{H}_\phi (k) \Psi_k
 \label{eq:charging_BdG}
\end{equation}
with the Nambu spinors $\Psi_k = (c_k , c_{-k}^\dagger)$, the charging energy $E_{\text{C}}$, and the bias charge $N_g$ (in a superconductor this arises due to a gate voltage).
The bosonic number operator $N$ and the phase mode $\phi$ are conjugate variables that satisfy $[\phi,N] = 2i$.

As is often the case when coupling a particle to a bath, $\phi$ interacts with every fermion mode~\cite{Caldeira:1981bf,Caldeira:1983jq,Grabert:1984gk,Dorsey:1985iw,Legget:1987di}; for our system this makes the coupling nonlocal. 
The system can be made local by including spatial fluctuations $\partial_x \phi$; in a superconducting system, neglecting these is justified provided that the length of the system is smaller than the Josephson penetration depth $\lambda_{\text{J}}$~\cite{Anderson:1964ce,TinkhamBook} and the characteristic length $L_H$~\cite{Ovchinnikov:2007jf,Ovchinnikov:2008hh} induced by a perpendicular magnetic field component~\cite{Stern:2019ji}.

Na\"ively, we could replace the fermionic Hamiltonian~\eqref{eq:fermionic_hamiltonian} by its ground state energy; this corresponds to the particle tunneling in the potential $V_\phi$. 
The resulting dynamics would be essentially described by a $0-\pi$ Cooper-pair box~\cite{Doucot:2002et,Ioffe:2002kr}.
Alluding to the superconductor analogy, we refer to energy scale characterizing $V_\phi$ as the Josephson energy, which for concreteness we define as $E_\text{J} \equiv (V_{\pi/2} - V_{0}) / 2$. 
The scenario we outlined in the Introduction is one where the fermionic bath is gapped at the potential minima. 
This implies $\omega_0\ll\Delta$ where $\omega_0\propto \sqrt{E_\text{C}E_\text{J}}$ is the Josephson plasma frequency: the characteristic energy scale for the oscillations of the particle in the potential minima.
We shall be interested in the regime where the tunneling between these minima can be read off from the splitting between the energies of the particle's ground and first excited states. 
We thus require this splitting to be much smaller than the level spacing $\omega_0$ characterizing each minima. 
In terms of the na\"ive potential tunneling picture, this corresponds to $E_{\text{J}}/E_{\text{C}} \gg 1$.

\section{Tunneling and Instantons}
\label{sec:instantons}

Our goal is to estimate the influence of the fermions' topology change on the amplitude for tunneling between $V_\phi$'s adjacent minima. 
Owing to the fermions being gapped near $V_\phi$'s minima, the particle simply experiences potential $V_\phi$ near the corresponding values of $\phi$. 
However, since the tunneling path involves a topology-changing gap closing, the picture of a particle tunneling in potential $V_\phi$ is qualitatively incomplete. 

To calculate the fermionic correction to the amplitude, one may consider taking a boson$\times$fermion factorized wave function ansatz near the potential minima and applying a generalization of linear combination of atomic orbitals (LCAO)~\cite{ashcroft1976solid} to this ansatz. 
While the exponential decaying overlap of the topologically distinct fermionic ground states already suggests a correction factor suppressing tunneling exponentially with the fermionic system size, LCAO is known to inaccurately capture tunneling exponents even in simple cases~\cite{harrell1980double,arzamasovs2017tight}. 
Hence, one might worry that elevating LCAO to our more intricate scenario might miss key features. 
We therefore use a field theoretical approach instead that can incorporate bosonic and fermionic features on the same footing. 
In such field theories, tunneling problems can be addressed via instantons~\cite{coleman_aspects_1985}, which will also be our method.
This is a semiclassical approach and hence requires working in the small tunnel splitting regime we are interested in~\footnote{The semiclassical nature of the small splitting regime can be seen from the tunneling exponent $\hbar^{-1}\int \sqrt{2m[V(x)-E)}]dx$ for a particle of mass $m$ and energy $E$ in a potential $V$.}.

\subsection{Path Integral}

Our starting point is the partition function corresponding to the Hamiltonian~\eqref{eq:charging_BdG}. 
It can be written as a path integral over fields defined on an interval of imaginary time $\tau \in [0, \beta)$:
\begin{equation}
  Z=\int\mathcal{D}\phi\mathcal{D}\boldsymbol{c}\mathcal{D}\bar{\boldsymbol{c}}\,e^{-S},
  \label{eq:partition_function}
\end{equation}
where $\boldsymbol{c}$, $\bar{\boldsymbol{c}}$ are Grassmann variables satisfying anti-periodic temporal boundary conditions.
The action $S = S_\phi + S_{\text{f}}$ is composed of a bosonic part
\begin{equation}
  S_\phi=\int_{0}^{\beta}d\tau\,\left[\frac{1}{2}\frac{1}{8E_\text{C}}\left(\partial_{\tau}\phi_{\tau}\right)^{2}+ i\frac{N_g}{2} (\partial_{\tau}\phi_\tau)\right]
  \label{eq:bosonic_action}
\end{equation}
and a fermionic part~\cite{altland_condensed_2010}
\begin{equation}
S_{\text{f}}=\frac{1}{2}\int_{0}^{\beta}d\tau\, \bar{\Psi}^{T}(\partial_{\tau}+\mathcal{H}_{\phi_\tau})\Psi,
\end{equation}
where $\Psi=\left(\boldsymbol{c},\bar{\boldsymbol{c}}\right)$ is the Nambu spinor and $\mathcal{H}_{\phi}$ is the fermionic Bogoliubov-de-Gennes (BdG) Hamiltonian~\eqref{eq:fermionic_hamiltonian}.
The phase $\phi_\tau$ appearing in the path integral is no longer compact and instead takes on any real value, subject to quasiperiodic boundary conditions $\phi_\beta = \phi_0 + 2\pi w$ which allow for nontrivial windings $w\in \mathbb{Z}$.
The $\beta\to\infty$ limit of $Z$ will give us information about the ground state of the system.

\subsection{The Instanton Gas}
\label{subsec:instanton_gas}

To gain intuition into the tunneling problem and a baseline to quantify the effect of the topology-changing fermions, we now consider a simpler problem, where we replace the fermionic Hamiltonian in Eq.~\eqref{eq:fermionic_hamiltonian} by its ground state energy $V_\phi$, a double-well potential with minima $V_0=V_\pi$ at $\phi = 0$ and $\phi = \pi$, respectively.
(This is exemplified by APBC in Fig.~\ref{fig:energies}(e) and can also be achieved for PBC and OBC by a suitable symmetrization to be described below.)
The resulting action $S_{\text{n}} = S_\phi + U_{\text{n}}$ is similar to Eq.~\eqref{eq:partition_function}, but since fermions are absent, the fermionic action $S_{\text{f}}$ has been replaced with what we dub the `na\"ive potential'
\begin{equation}
  U_{\text{n}}[\phi_\tau] = \int_{0}^{\beta}d\tau\, V_{\phi_\tau}.
  \label{eq:naive_potential}
\end{equation}

The dominant contributions to this path integral come from $\phi_\tau$ configurations that minimize $S_{\text{n}}$, the most significant being $\phi_\tau = 0$ and $\phi_\tau = \pi$, which stay at a minimum of the potential throughout.
The action also has other subsidiary minima that contribute to $Z$ at next-leading order, arising from extremal configurations of $\phi_\tau$ which solve $\delta S_{\text{n}} = 0$.
Looking at the Euler-Lagrange equations resulting from $S_{\text{n}}$, the dynamics of $\phi$ (thought of as a position coordinate) are equivalent to a classical particle moving in a potential landscape $-V_\phi$~\cite{altland_condensed_2010}.
Thus in addition to staying at the top of the potential, the particle can move down from one extremum and up to another, where it can spend an arbitrary amount of time before returning, if the minima of $V_\phi$ are symmetric.
The $\phi_\tau$ solutions that connect the minima are called \emph{instantons} (or \emph{anti-instantons} when moving in the opposite direction)~\cite{coleman_aspects_1985}.
Their name refers to them being well-localized in (imaginary) time: their width (i.e., the time spent between the potential extrema) is of the order $1/\omega_0$, the harmonic oscillator frequency of the two wells, assumed to be the same.
Each classical instanton has action $S^{\star}_{\text{n}}$, and these can be chained together to form approximate solutions of the classical equations of motion.
To capture the effect of tunneling on the low-energy spectrum, we sum over all such solutions in what is known as the \emph{instanton gas} summation~\cite{coleman_aspects_1985}.

Following \citeauthor{coleman_aspects_1985}'s  calculation~\cite{coleman_aspects_1985}, the symmetry of the minima $V_0 = V_\pi$ allows $q$ instantons and $\bar{q}$ anti-instantons to appear at any time and in any order (provided one starts and ends in a minimum of the same type).
Summing over all winding numbers $w\in\mathbb{Z}$, the leading terms in the partition function are given by
\begin{equation}
  Z \propto e^{-\beta \left( V_0 + \frac{\omega_0}{2} \right)} \sum_w \sum_{q,\bar{q}=0}^{\infty} \delta_{w,\frac{q-\bar{q}}{2}} \frac{\left( \beta K e^{-S^{\star}_{\text{n}}} \right)^{q + \bar{q}}  e^{-i \frac{\pi}{2} N_g \left(q - \bar{q}\right)}}{q! \, \bar{q}!}.
  \label{eq:instanton_gas}
\end{equation}
$K$ is a fluctuation factor associated with each instanton, whose value shall not concern us.
Splitting up the $\sum_{q/\bar{q}}$ summations into even and odd contributions, we see that the partition function is given by
\begin{equation}
  Z \propto e^{-\beta \left( V_0 + \frac{\omega_0}{2} \right)} \cosh{\left[ 2\beta K e^{-S^{\star}_{\text{n}}} \cos{\left( \frac{\pi}{2} N_g \right)} \right]}.
\end{equation}
By comparing this to the form $Z = \Tr{\left[ \sum_{n} e^{-\beta E_n} \ket{n} \bra{n} \right]}$, we deduce that the low-energy spectrum is given by
\begin{equation}
  E_{\pm}(N_g) = V_0 + \frac{\omega_0}{2} \pm 2K e^{-S^{\star}_{\text{n}}} \cos{\left( \frac{\pi}{2} N_g \right)},
  \label{eq:tight_binding_spectrum}
\end{equation}
which is the familiar tight-binding dispersion with tunneling amplitude $ t_{0\to\pi} = 2 K e^{-S^{\star}_{\text{n}}}$ between states at $\phi = 2m\pi$ and $\phi = (2m+1)\pi$ that otherwise have equal energy.
The action appearing in the exponent consists of equal kinetic and potential energy parts and is given by
\begin{equation}
  S^{\star}_{\text{n}} = \frac{1}{2\sqrt{E_{\text{C}}}} \int_{0}^{\pi} d\phi \sqrt{V_\phi - V_0},
  \label{eq:classical_action}
\end{equation}
which is the source of the exponential suppression of charge noise with $E_\text{J} / E_\text{C}$ in transmons~\cite{Koch:2007gz,Schreier:2008gs}.

A key assumption behind the instanton gas summation is that the gas is dilute~\cite{coleman_aspects_1985}. 
The instanton density is $\propto \exp(-S^{\star}_{\text{n}})$~\cite{coleman_aspects_1985}, hence the diluteness assumption amounts to requiring small tunnel splitting, placing us in the regime we are interested in.

Note that to be able to read out the $0 \to \pi$ tunneling amplitude from the spectrum, we required the minima to be equal. 
Had this not been the case, supposing instead that $V_0 < V_\pi$, the classical solutions would more closely resemble a sequence of $0 \to 2\pi$ instantons separated by large imaginary time durations, rather than $0 \to \pi$ instantons~\cite{rodriguez-mota_revisiting_2019}.
In this limit we would get a ground state energy that disperses as $\cos(\pi N_g)$, i.e., with halved $N_g$ periodicity, corresponding to the $2\pi$-periodicity of the potential encoding Cooper pair tunneling (instead of the electron quartet tunneling encoded by $\pi$ periodicity~\cite{Doucot:2002et,Ioffe:2002kr}).
We mention this because the ground state energy of the double-well Kitaev model, as written in Eq.~\eqref{eq:fermionic_hamiltonian}, is symmetric only for APBC, as seen from Fig.~\ref{fig:energies}.
With PBC, although the minima have equal energies, their ground state parities are different due to the gap closing, and hence the minima are completely decoupled for fixed parity.
OBC has an asymmetric ground state profile even without fixing parity because the gapped bulk modes are replaced by zero-energy Majorana end modes in the topological phase, which do not contribute to the ground state energy.
For PBC and OBC, we therefore have to symmetrize the model in order for the magnitude of the $0 \to \pi$ tunneling to be visible in the low-energy spectrum.
Due to its closer link to our inspiring topological superconducting systems~
\cite{pientka_topological_2017,Hell:2017fn,Fornieri:2019hz,Ren:2019ew} and features that Majorana end modes may present, of these two cases we mainly focus on OBC, while we include APBC for its relative technical simplicity.  

\subsection{Integrating Out Fermions}
\label{subsec:integrating_out_fermions}

Our goal is to capture the modification of $t_{0\to\pi}$ due to the topology-changing fermions.
To compute this, we now return to the full many-body path integral in Eq.~\eqref{eq:partition_function}.
The fermionic Lagrangian is bilinear in the fermionic fields and so we can perform a Gaussian integral to obtain~\cite{zinn-justin_quantum_2021}
\begin{equation}
  \int \mathcal{D}\boldsymbol{c} \mathcal{D}\bar{\boldsymbol{c}}\, e^{-S_{\text{f}}} = \Pf{\left[ \partial_\tau + \mathcal{H}_{\phi_\tau}^{M} \right]},
\end{equation}
where $\mathcal{H}_{\phi}^{M} = - [\mathcal{H}_{\phi}^{M} ]^T = W \mathcal{H}_{\phi} W^{\dagger}$ is the Hamiltonian written in the Majorana basis with $W=\frac{1}{\sqrt{2}} \begin{psmallmatrix} 1 & 1 \\ -i & i \end{psmallmatrix}$.
Thus by integrating out the fermions we have obtained a partition function expressed as a path integral of the phase only
\begin{equation}\label{eq:ZPfaff}
  Z = \int\mathcal{D}\phi \,e^{-S_\phi} \Pf{[ \partial_\tau + \mathcal{H}_{\phi_\tau}^M ]} \big/ \Pf{[ \partial_\tau + \mathcal{H}_{0\text{i}}^M ]},
\end{equation}
albeit one with a complicated (temporally nonlocal) action.
Because of the continuous $\partial_\tau$ term, the Pfaffian needs to be regularized and so we divide by the Pfaffian for the static phase profile $\phi_{0\text{i}} = 0$ without instantons.
The suppression of tunneling due to topology-changing fermions originates in the deviation of this Pfaffian ratio from the na\"ive potential, a point we will further elucidate in Sec.~\ref{sec:fermionic_factor} (where we will also find that our regularization is analogous to an offset sending $V_0\rightarrow 0$).

Since we are interested only in the magnitude of this fermionic suppression, it will be simpler work with the determinant. 
In terms of this,
\begin{equation}
   \sqrt{\det{\left[ \partial_\tau + \mathcal{H}_{\phi_\tau} \right]}} =  \Pf{[ \partial_\tau + \mathcal{H}_{\phi_\tau}^M ]} 
  \label{eq:determinant_is_pfaffian}
\end{equation}
up to a sign that plays no role in our considerations~\footnote{For the chirally symmetric system we consider, we always have a real quantity $\det{[ \partial_\tau + \mathcal{H}_{\phi_\tau} ]} \big/ \det{[ \partial_\tau + \mathcal{H}_{0\text{i}} ]} > 0$ for all $\phi_\tau$, so the sign of the corresponding Pfaffian never changes.}.
This fermionic factor defines what we refer to as the `fermionic potential'
\begin{equation}
  U_{\text{f}}{\left[ \phi_\tau \right]} \equiv - \frac{1}{2}\log{\det{[ \partial_\tau + \mathcal{H}_{\phi_\tau} ]}},
\end{equation}
although as it stands only $U_{\text{f}}{\left[ \phi_\tau \right]}-U_{\text{f}}{\left[ \phi_{0i} \right]}$ corresponding to the Pfaffian ratio is well defined. 
Unless stated otherwise (cf.\ Sec.~\ref{sec:fermionic_factor}), henceforth we consider this difference and compare with the na\"ive case with a similarly subtracted na\"ive potential (this subtraction is just an inconsequential energy offset in the na\"ive case).
Crucially, the relevant phase profiles $\phi_\tau$ contributing to the path integral still resemble those of the instanton gas, so our partition function can be expanded in the same way as Eq.~\eqref{eq:instanton_gas} but with modified action.

\section{Tunneling Suppression via Topology in $d+1$ Dimensions}
\label{subsec:low_energy_analogy}

We next explain how the instantons connecting minima where $\mathcal{H}_{\phi_\tau}$ has distinct topology can be linked to topologically protected $d$-dimensional gapless boundary modes a topological Hamiltonian in $d+1$ dimensions. 
At the core of this correspondence is imaginary time supplying an extra dimension that, in a manner akin to reversing dimensional reduction~\cite{kitaev2009periodic,Ryu_2010,QiZhang_RevModPhys.83.1057,bernevig2013topological}, allows one to climb a step higher in a dimensional hierarchy.

By Eq.~\eqref{eq:determinant_is_pfaffian}, we require the product of all eigenvalues of the (non-Hermitian) kernel $\mathcal{L}(\tau) = \partial_\tau + \mathcal{H}_{\phi_\tau}$.
When $\mathcal{H}_{\phi_\tau}$ enjoys a chiral symmetry, $\left\{ \Gamma,\, \mathcal{H}_{\phi}  \right\} = 0 $ with $\Gamma$ a gamma matrix (i.e., a Pauli matrix or its Hermitian higher-dimensional generalization~\cite{de1986field}), then $\tilde{\mathcal{H}}(\tau) = i\Gamma \mathcal{L}(\tau)$ is a Hermitian operator. 
[Note that $\det{\tilde{\mathcal{H}}(\tau)} = \det{\mathcal{L}(\tau)}$ since $\det{i\Gamma} = 1$.]
If chiral symmetry is absent, it can be introduced by doubling, i.e., considering 
$\mathcal{L}'(\tau) = \partial_\tau + \mathcal{H}_{\phi_\tau}\otimes \sigma_a$ with Pauli matrix  $\sigma_a$ (and taking another square root of the corresponding determinant to recover the Pfaffian, as done in the $d=0$ example of Ref.~\onlinecite{pekker_suppression_2013}). 
Now $\tilde{\mathcal{H}}(\tau) = i\Gamma \mathcal{L}'(\tau)$, with $\Gamma = \openone\otimes \sigma_b$ ($b\neq a$), which is again Hermitian. 
The operator $\tilde{\mathcal{H}}(\tau)$ can be interpreted as a Hamiltonian in $d+1$ dimensions. 
The steps leading to $\tilde{\mathcal{H}}(\tau)$, including the doubling in the non-chiral case, parallel closely (the reversal of) features in dimensional reduction procedures for topological insulators and superconductors~\cite{Ryu_2010}. 

Topologically protected gapless interface states at instanton locations are guaranteed to arise because, at low energies, the topological transition (with $\phi_\tau$) of  $\mathcal{H}_{\phi_\tau}$ reduces to a mass inversion of a $d$-dimensional Dirac Hamiltonian which, in turn, becomes a $\tau$-dependent mass kink for $\tilde{\mathcal{H}}(\tau)$ in $d+1$ dimensions. 
Such mass kinks, by a generalization of the Jackiw-Rebbi mechanism~\cite{Jackiw:1976ky}, bind $d$-dimensional topologically protected gapless modes, a key feature underlying topological insulators' and superconductors' robust boundary modes~\cite{Schnyder_PhysRevB.78.195125,kitaev2009periodic,Ryu_2010}. 

\begin{figure}
  \includegraphics[width=\linewidth]{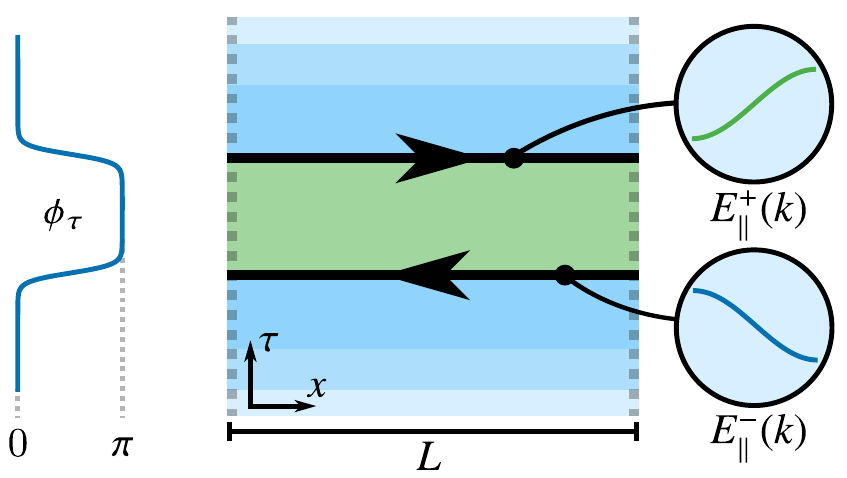}
  \caption{A cartoon showing the edge modes of $\tilde{\mathcal{H}}(k,\tau)$ for an instanton-anti-instanton configuration (left) with twisted boundary conditions.
  On the right are the dispersions $E_{\parallel}^{\pm}(k)$ of the two edge modes, showing they have different chirality.}
  \label{fig:APBC_phase_transition}
\end{figure}

For our toy model Eq.~\eqref{eq:fermionic_hamiltonian}, $\{ \sigma_1,\, \mathcal{H}_{\phi} \} = 0 $ and thus 
\begin{align}
  \tilde{\mathcal{H}}(\tau) = i\sigma_1 \mathcal{L}(\tau)
  =& \frac{\Delta}{2}\cos{\phi_{\tau}} \left[\left(1+\cos{k}\right)\sigma_{2}-\sin{k}\sigma_{3}\right] \nonumber \\
  &+\frac{\Delta}{2}\left[\left(1-\cos{k}\right)\sigma_{2}+\sin{k}\sigma_{3}\right]+i\partial_{\tau}\sigma_{1}
\end{align}
is Hermitian. 
It is a $d=2$ class D superconductor Hamiltonian.
Each instanton, i.e., a phase slip of $\phi$ by $\pi$, corresponds to a topological transition of $\tilde{\mathcal{H}}(\tau)$; the corresponding interface states are shown schematically in Figs.~\ref{fig:APBC_phase_transition} and \ref{fig:OBC_phase_transition}.

\subsection{Sharp Instanton Limit}
\label{subsec:sharpinst}

While the shape of $\phi_\tau$ and the fermionic spectrum are interdependent and hence solving for them is a nontrivial problem, certain limiting cases for $\phi_\tau$ allow for tractable examples that illuminate generic topological features dictating the behavior of the Pfaffian ratio. 
We next focus on such a case, specifically on the ``sharp instanton limit'' of instantons with vanishingly short width. 
Although, since the instanton width is set by $1/\omega_0$, such sharp instantons are beyond the $\omega_0\ll \Delta$ regime, they are not only analytically tractable, but (as we shall justify in Sec.~\ref{subsec:strategy}) they also correspond to the biggest discrepancy between the fermionic and na\"ive potentials, and hence will allow us to bound the fermionic suppression of $t_{0\to\pi}$.

A key simplification of the sharp instanton limit is that for such instantons there is no time spent away from the minima to accumulate potential contributions to the na\"ive action and so $U_{\text{n}}{\left[ \phi_{2m\text{i}} \right]} = U_{\text{n}}{\left[ \phi_{0\text{i}} \right]}$ for a profile $\phi_{2m\text{i}}$ with $2m$ sharp instantons. 
For the fermionic potential, however, we will show that $U_{\text{f}}{\left[ \phi_{2m\text{i}} \right]} \gg U_{\text{f}}{\left[ \phi_{0\text{i}} \right]}$.
Evaluating the Pfaffian ratio for a profile $\phi_{2m\text{i}}$ amounts to comparing the energies of $\tilde{\mathcal{H}}_{2m\text{i}}(\tau)$ with those of a static Hamiltonian $\tilde{\mathcal{H}}_{0\text{i}}(\tau)$, where $\tilde{\mathcal{H}}_{n\text{i}}(\tau)$ is $\tilde{\mathcal{H}}(\tau)$ on the background of $n$ sharp instantons.
Since the spectrum of $\tilde{\mathcal{H}}(\tau)$ is qualitatively different for OBC versus APBC, we discuss each case separately.

\subsubsection{APBC}

Antiperiodic boundary conditions are the simplest to deal with:
Translational invariance means we can stay in momentum space along the spatial direction.
Each instanton changes the sign of $\cos{\phi_\tau}$ and corresponds to a topological transition of $\tilde{\mathcal{H}}(k, \tau)$ that binds low-energy chiral modes with dispersion
\begin{equation}
  E_{\parallel}^{\pm}(k) = \pm\Delta \sin{(k / 2)}
  \label{eq:chiral_mode_dispersion}
\end{equation}
to the interface (derived in Appendix~\ref{sec:Jackiw-Rebbi}), where chirality depends on the direction of the sign change of $\cos{\phi_\tau}$.
To find the contribution of each instanton to $\det{\tilde{\mathcal{H}}(\tau)}$, we take the product of the energies of all sub-gap states sampled by momenta $k_n = (2n+1)\pi / L$ consistent with APBC for $L$ sites.
All these states would otherwise be at the gap energy $\Delta$, so the determinant ratio for a configuration with $2m$ phase transitions is given, up to an inconsequential sign, by 
\begin{equation}
\label{eq:2miAPBC}
  \frac{\det{\tilde{\mathcal{H}}_{2m\text{i}}(\tau)}}{\det{\tilde{\mathcal{H}}_{0\text{i}}(\tau)}} = \left[ \prod_{n=0}^{L-1} \sin{(k_n / 2)} \right]^{2m} = \left[ 2^{-(L-1)} \right]^{2m},
\end{equation}
since other supra-gap states remain unchanged.
The last equality makes use of a trigonometric identity~\footnote{This is a specific case of the general identity $\sin{(Lx)} = 2^{L-1} \prod_{k=0}^{L-1} \sin{(x + k\pi/L)}$ which follows from writing the $L$ roots of unity as $z^L-1 = \prod_{k=0}^{L-1} ( z - e^{-i2\pi k / L} )$. Dividing by the first term and taking $x \to 0$ also gives $L = 2^{L-1} \prod_{k=1}^{L-1} \sin{(k\pi/L)}$, which can be used for PBC.}, but one expects an $e^{-\alpha L}$ dependence for edge mode dispersions of any shape since the logarithm of the product can be approximated by an integral in the large $L$ limit
\footnote{For a positive function $f(k)$ we have $\log{\prod_{n=0}^{L-1} f(k_n)}=\sum_{n=0}^{L-1} \log{f(k_n)} \to ([L-1]/2\pi)\int_0^{2\pi} dk\,f(k)$ for $L \gg 1$.}.

When instantons are very close, the exponentially localized edge modes with opposite chirality can overlap and gap out, but as instantons separate, this small exponential splitting is quickly dwarfed by $E_{\parallel}^{+}(\pi/L)$.
Thus, the dilute instanton gas approximation~\cite{coleman_aspects_1985} remains valid.
(In fact, the approximation is even better justified than in the na\"ive case since the instanton density is exponentially suppressed in the instanton action~\cite{coleman_aspects_1985} so the gas is further rarefied in the presence of fermions due to the increased instanton action.)

Generalizing the calculation of Eq.~\eqref{eq:tight_binding_spectrum} to the case with fermions, the tunnel amplitude is still set by the single-instanton action, which is half of that of the two-instanton case, the minimal configuration allowed by the $\phi_0\equiv \phi_\beta$ temporal boundary conditions imposed by the partition function. 
(While $\phi\equiv \phi+2\pi$ due to charge quantization, $\phi \not\equiv \phi +\pi$ notwithstanding $V_0=V_\pi$.)
Although the na\"ive potential $U_{\text{n}}\left[\phi_{2\text{i}}\right] - U_{\text{n}}\left[\phi_{0\text{i}}\right] \to 0 $ vanishes for sharp instantons $\phi_{2\text{i}}$ (where the subtraction of $U_{\text{n}}\left[\phi_{0\text{i}}\right]$ follows the naive potential limit of the Pfaffian ratio, cf.\ Sec.~\ref{sec:fermionic_factor}), 
the topologically guaranteed chiral modes of $ \tilde{\mathcal{H}}_{2\text{i}}(\tau) $ mean that the fermionic potential approaches a lower bound $U_{\text{f}}\left[\phi_{2\text{i}}\right] - U_{\text{f}}\left[\phi_{0\text{i}}\right] \to (L-1)\log{2}$ upon reducing the instanton width.

This sharp instanton limit is the regime with the strongest suppression of tunneling due to fermions (cf.\ Sec.~\ref{subsec:strategy}), and so by modifying the instanton action appearing in the spectrum~\eqref{eq:tight_binding_spectrum}, we can bound by how much the na\"ive tunneling amplitude $t^{\text{(n)}}_{0\to\pi}$ can be modified by fermions. For APBC, Eq.~\eqref{eq:2miAPBC} implies
\begin{equation}
  t^{\text{(f)}}_{0\to\pi} \geq e^{-\frac{1}{2}(L-1)\log{2}}\, t^{\text{(n)}}_{0\to\pi}.
  \label{eq:bandwidth_suppression_sharp}
\end{equation}
Thus the tunneling is exponentially suppressed as a function of system size.
Since it derives entirely from the topological boundary modes of $ \tilde{\mathcal{H}}_{2\text{i}}(\tau) $, this bound on the scaling exponent is purely due to the topological inequivalence of the two ground states, and is unrelated to the trivial scaling of $V_\phi$ with $L$.
(Any information about the energy scale $\Delta$ was lost when taking the ratio of energies, but this is unique to the sharp instanton limit since we will later see that the scale of the potential influences the instanton timescale and hence the fermionic factor.)

\subsubsection{OBC}
\label{subsubsec:low_energy_analogy_obc}

\begin{figure}
  \includegraphics[width=\linewidth]{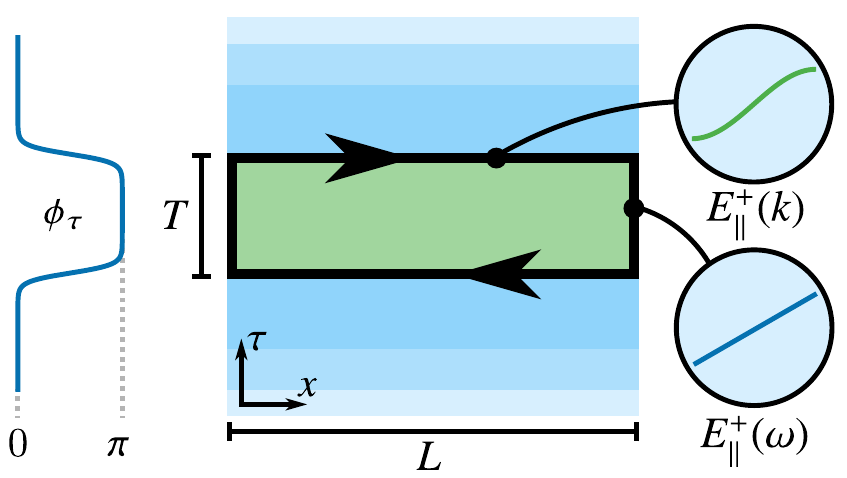}
  \caption{A cartoon showing the edge mode for an instanton-anti-instanton configuration with OBC, separated by imaginary time $T$.
  The chiral mode runs along the entire boundary of the topological region.
  Shown on the right are the dispersions along different sections of the perimeter.}
  \label{fig:OBC_phase_transition}
\end{figure}

Open boundary conditions are technically more difficult to treat because we cannot work in momentum space.
Recall that the topological limit of the 1D Kitaev chain has zero-energy Majorana end modes~\cite{Kitaev:2001gb}.
When adding the $i\partial_\tau \sigma_1$ term, these Majorana modes gain a linear dispersion in the temporal direction, given by
\begin{equation}
  E_{\parallel}^{\pm}(\omega) = \pm \omega.
\end{equation}
This means that for two instantons separated by imaginary time $T$, instead of having counterpropagating chiral edge modes in a ribbon geometry as for APBC, these edge modes run all the way around the perimeter of the 2D $(T \times L)$-sized topological region, shown in Fig.~\ref{fig:OBC_phase_transition}.
The differing dispersions in each direction mean that the frequency $\omega$ and wave vector $k$ on the different sections are related through the energy matching condition
\begin{equation}
  E_{\parallel}^{+}(\omega) = E_{\parallel}^{+}(k).
  \label{eq:energy_matching}
\end{equation}
As a fermion traverses this perimeter, it undergoes one whole rotation and must acquire a phase $e^{i\pi}$, which gives a quantization condition
\begin{equation}
 2T\omega + 2Lk = (2n+1)\pi,
  \label{eq:chiral_quantization}
\end{equation}
with $n \in \mathbb{Z}$ that can be solved simultaneously with Eq.~\eqref{eq:energy_matching} to obtain the quantized energies of the chiral modes.
We then follow the same principle of taking the product of these energies as a fraction of the gap.

Although the quantization condition lacks algebraic solutions, we numerically find that Eq.~\eqref{eq:chiral_quantization} leads to a product [generalizing Eq.~\eqref{eq:2miAPBC}] that depends exponentially on $T$.
Intuitively, this is a consequence of the asymmetry in the ground state for OBC: 
Recall from Fig.~\ref{fig:energies} that for OBC the ground state minima are offset by $\Delta$, which means that the na\"ive action would accumulate a contribution $\int d\tau \, V_{\phi_\tau} = T \Delta$ when $\phi_\tau$ spends a duration $T$ in the higher well.
The result for multiple pairs of instantons follows by summing over all pair separations $\tau_{2j} - \tau_{2j-1}$ which govern the sizes of topological regions.
The presence of $e^{-(\tau_{2j} - \tau_{2j-1})\Delta}$ terms prevents us from plugging our result into Eq.~\eqref{eq:instanton_gas} as a modified instanton action because the integral over instanton locations $\{ \tau_i\}$ (a key step leading to Eq.~\eqref{eq:instanton_gas}, c.f. Ref.~\cite{coleman_aspects_1985}) will be fundamentally different---the instantons are now \emph{interacting}.
However, as mentioned in Sec~\ref{subsec:instanton_gas}, asymmetric wells do not tell us about $0\to\pi$ tunneling, so we must symmetrize the wells. We therefore change the gap on the trivial side to $\Delta' = \Delta (1 - 1/L)$. 
One intuitively expects that having thus symmetrized the wells, i.e., removed the $\int d\tau \, V_{\phi_\tau} = T \Delta$ contribution, we have eliminated the instanton interactions, which allows us to interpret what remains as a modified instanton action. 
This will indeed turn out to be the case, however it requires more careful justification: by symmetrizing the wells we introduced a gap asymmetry and this requires us to consider more than just the chiral edge modes (which themselves are also modified).

The first modification is that when changing $\Delta \to \Delta'$ in the trivial phase, the dispersion of the chiral modes along $x$ is replaced by $E_{\parallel}^{\prime \pm}(k)$, as we detail in Appendix~\ref{sec:Jackiw-Rebbi}, which changes the energy matching equation~\eqref{eq:energy_matching}.
Letting $k_n^{\star}$ denote the quantized momenta of states running around the edge of length $2T + 2L$, we numerically find that \begin{equation}
  \prod_{n} \frac{E_{\parallel}^{\prime+}{(k_n^{\star})}}{\Delta'} \approx \exp{\left[-2(L-1)\log{2} - 4T \Delta' / \pi \right]}.
  \label{eq:obc_sub-gap}
\end{equation}

For OBC, energies above the gap are also modified by the existence of a topological region.
There are two kinds of supra-gap states that are modified.
The first kind is the set of states that are localized at the ends of the chain, but propagate along the temporal direction.
In the trivial gapped region, these end states have dispersion
\begin{equation}
  E_{\Delta'}^{\pm}(\tilde{\omega}) = \pm\sqrt{\tilde{\omega}^2 + \Delta'^2}
\end{equation}
that must match the energy $E_{\parallel}^{\pm}(\omega)$ in the topological region.
This time, the quantization condition comes from the APBC for fermions in the path integral:
\begin{equation}
  (\beta - T) \tilde{\omega} + T \omega = (2n+1)\pi.
  \label{eq:apbc_quantization}
\end{equation}
The supra-gap energies in the absence of any instantons are recovered by setting $T=0$ in the above equation.
Denoting the quantized frequencies by $\tilde{\omega}_n^{\star}$ and counting both positive and negative energies, as $\beta \to \infty$ the relevant ratio tends to
\begin{equation}
  \left[ \prod_{n} \frac{E_{\Delta'}^{+}(\tilde{\omega}_n^{\star})}{\sqrt{(2n+1)^2 \pi^2/\beta^2 + \Delta'^2}} \right]^2 = \exp{[- 2T\Delta' (1 - 2/\pi)]}.
  \label{eq:obc_supra-gap}
\end{equation}
The other kind of supra-gap states describes those in the bulk of the superconducting chain, whose energies are affected by the presence of instantons only because we have $\Delta' \neq \Delta$, but not for topological reasons. 
These are solved through a quantization condition similar to Eq.~\eqref{eq:apbc_quantization}, but this time one must match $E_{\Delta'}^{\pm}(\tilde{\omega})$ with $E_{\Delta}^{\pm}(\omega) = \pm \sqrt{\omega^2 + \Delta^2}$.
Each of these states are $(L-1)$-fold degenerate because each can be localized on any of the $2L - 2$ Majoranas not on the ends of the chain.
Since $\Delta > \Delta'$, these energies \emph{increase} when more time is spent in the topological phase (with states in the band $ \Delta' \leq | E_{\Delta'}^{\pm}(\tilde{\omega})| \leq \Delta$ pushed to be above $\Delta$) and their combined effect will be to cancel the exponential decay with instanton separation that we have seen in Eqs.~\eqref{eq:obc_sub-gap} and~\eqref{eq:obc_supra-gap}.
In the same $\beta \to \infty$ limit, we now have
\begin{equation}
  \left[ \prod_{n} \frac{E_{\Delta}^{+}(\omega_n^{\star})}{\sqrt{(2n+1)^2 \pi^2/\beta^2 + \Delta'^2}} \right]^{2(L-1)} = \exp{[2T\Delta']}.
\end{equation}

Combining all these contributions, the determinant ratio for two sharp instantons separated by $T$ is 
\begin{equation}
  \frac{\det{\tilde{\mathcal{H}}_{2\text{i}}(\tau)}}{\det{\tilde{\mathcal{H}}_{0\text{i}}(\tau)}} \approx \exp{[-2(L-1)\log{2}]}.
  \label{eq:suppression_edge_modes}
\end{equation}
Thus, after symmetrization, one finds the same bound on the suppression due to fermions for OBC as in Eq.~\eqref{eq:bandwidth_suppression_sharp} for APBC.

\section{Tunneling Suppression from a Variational Approach}
\label{sec:variational}

Although the argument based on topological edge modes quickly gave us an upper bound on the suppression due to fermions, it cannot easily be extended to give full quantitative results.
The problem is that typical instantons in the gas are not perfectly sharp, and instead have a finite timescale.
Despite the spectrum of the topological edge modes being independent of instanton shape, a finite instanton timescale leads to other nontopological bound states at the phase transition whose energies \emph{do} depend on instanton shape~\cite{charmchi_complete_2014}.
Analytic results for the full spectrum of a generic phase profile do not exist, and an approximate spectrum would not suffice because estimating the tunneling suppression relies on the precise difference between the fermionic determinant and its na\"ive equivalent.

\subsection{Fermionic Factor as a Generalization of the Ground State Potential}
\label{sec:fermionic_factor}

We now describe an exact approach to calculating the fermionic potential. 
This approach works directly with the kernel underlying the Pfaffian, without requiring converting to a Hermitian matrix and hence chiral symmetry. 
It will also give an interpretation of the fermionic determinant by linking it to the one-dimensional potential generated by the ground state of the BdG system.

We start by writing
\begin{equation}
  \det{\left[ \partial_\tau + \mathcal{H}_{\phi_\tau} \right]} = \prod_{nm} \lambda_{nm},
\end{equation}
where $\lambda_{nm}$ are the eigenvalues of the differential equation
\begin{equation}
  \left[ \partial_\tau + \mathcal{H}_{\phi_{\tau}} \right] \psi_{nm}(\tau) = \lambda_{nm} \psi_{nm}(\tau),
\end{equation}
which has eigenfunctions of the form
\begin{equation}
  \psi_{nm}(\tau) = \mathcal{T}\exp\left[\int_{0}^{\tau}d\tau^{\prime}\,\left( \lambda_{nm}-\mathcal{H}_{\phi_{\tau^{\prime}}} \right) \right] \psi_{nm}(0).
\end{equation}
Time-ordering (with later times appearing on the left) is required because the BdG Hamiltonian does not generally commute at different times, $\left[ \mathcal{H}_{\phi_\tau},\, \mathcal{H}_{\phi_{\tau^\prime}} \right] \neq 0 $.
Temporal APBC for fermions $\psi_{nm}(\beta) = -\psi_{nm}(0)$ fixes the eigenvalues to be
\begin{align}
  \lambda_{nm} &= i\omega_n + \frac{1}{\beta}\log{\left\{ \mathcal{T}\exp\left[-\int_{0}^{\beta}d\tau\, \mathcal{H}_{\phi_{\tau}} \right] \right\}}_m  \label{eq:eigenvalues_of_time_ordered} \\
  &\equiv i\omega_n - \left\{ \mathcal{H}_{\text{eff}}[\phi_\tau] \right\}_m,
\end{align}
where $\omega_n = (2n+1)\pi/\beta$ with $n \in \mathbb{Z}$ are the Matsubara frequencies and $\left\{ \bullet \right\}_m$ denotes the $m$th eigenvalue of an operator we denote as the effective Hamiltonian $\mathcal{H}_{\text{eff}}[\phi_\tau]$ (emphasizing that it depends on the entire $\phi_\tau$ profile).
The spectrum of $\mathcal{H}_{\text{eff}}[\phi_\tau]$ inherits PH symmetry.

When taking the product of these eigenvalues, we may use the Weierstrass factorization theorem to rewrite the determinant as~\cite{Dashen:1975ie}
\begin{equation}
  \det{\left[ \partial_\tau + \mathcal{H}_{\phi_\tau} \right]} =  \prod_{m} \left( \prod_n i\omega_n \right) \cosh{\left[\frac{\beta}{2}\left\{ \mathcal{H}_{\text{eff}}[\phi_\tau] \right\}_m \right]}.
  \label{eq:determinant_product}
\end{equation}
Upon taking the ratio, the normalization-dependent prefactor drops out to give
\begin{equation}
  \frac{\det{[ \partial_\tau + \mathcal{H}_{\phi_\tau}]}}{\det{[ \partial_\tau + \mathcal{H}_{0\text{i}} ]}} = \prod_m \frac{\cosh{[(\beta / 2) \left\{ \mathcal{H}_{\text{eff}}[\phi_\tau] \right\}_m]}}{\cosh{[(\beta / 2) \left\{ \mathcal{H}_{\text{eff}}[\phi_{0\text{i}}] \right\}_m]}},
  \label{eq:fermionic_cosh}
\end{equation}
whose numerator is the partition function for a BdG Hamiltonian $\mathcal{H}_{\text{eff}}[\phi_\tau]$.

To make the link to the na\"ive potential, consider a case where $\mathcal{H}_{\phi_\tau}$ commutes at all times, such that $\mathcal{H}_{\text{eff}}[\phi_\tau] = \frac{1}{\beta} \int_0^{\beta} d\tau\, \mathcal{H}_{\phi_\tau}$
and hence $\left\{ \mathcal{H}_{\text{eff}}[\phi_\tau] \right\}_m = I_m[\phi_\tau] = \frac{1}{\beta} \int_0^{\beta} d\tau \varepsilon_{\phi_\tau, m}$ with $\varepsilon_{\phi_\tau, m}$ the instantaneous single particle energies of $\mathcal{H}_{\phi_\tau}$ (taken to evolve continuously with $\phi_\tau$ through any level crossing). 
In the sense of $I_m[\phi_\tau]$, both the APBC and OBC systems are gapped provided $\phi_\tau$ spends significant time near $\phi = 0$. 
Therefore, when ignoring the evolution of eigenstates with superconducting phase, the fermionic factor tends (upon taking $\beta$ much larger than the inverse of the $I_m$ gap) to the action of a potential that is the ground state energy. 
By the same logic, by noting $I_m[\phi_{0\text{i}}]=\varepsilon_{\phi=0,m}$, the denominator in Eq.~\eqref{eq:fermionic_cosh} can be seen to subtract $\varepsilon_{\phi=0,m}$ from each $\varepsilon_{\phi_\tau, m}$, thus supplying an offset setting the minimum value of this potential to zero.

When the eigenstates of the BdG Hamiltonian evolve as a function of phase, Eq.~\eqref{eq:fermionic_cosh} is viewed as the generalization of the ground state potential action and the spectrum of the time-ordered quantity $\mathcal{H}_{\text{eff}}[\phi_\tau]$ must be evaluated properly.
The importance of eigenstate evolution is also clear if we diagonalize the BdG Hamiltonian in the path integral from the outset as $\mathcal{H}_{\phi} = X_{\phi} [ \oplus_{m}^{} \varepsilon_{\phi, m}^+ \sigma_3 ] X_{\phi}^{\dagger}$.
Then, the fermionic determinant is replaced by~\footnote{This equivalent form is generally less convenient for numerical calculations for the same reasons that Wilson loops are often more convenient than Berry phase integrals.}
\begin{equation}
  \det{\left[\partial_\tau + \mathcal{H}_{\phi_\tau}\right]} \to \det{\left[\partial_\tau + \oplus_{m}^{} \varepsilon_{\phi_\tau, m}^+ \sigma_3 + X_{\phi_\tau}^\dagger \partial_\tau X^{}_{\phi_\tau} \right]},
  \label{eq:Lagrangian_alternative}
\end{equation}
where we see the last term, i.e., the eigenstate evolution, being responsible for the deviation from $\left\{ \mathcal{H}_{\text{eff}}[\phi_\tau] \right\}_m=I_m[\phi_\tau]$, i.e., from the na\"ive case.

These considerations, in particular the cancellation in Eq.~\eqref{eq:fermionic_cosh}, also show how one can define an unsubtracted variant of the fermionic potential: the functional 
\begin{equation}
\label{eq:standalone}
\tilde{U}_{\text{f}}[\phi_\tau]=-\frac{1}{2}\sum_m\log\cosh{[(\beta / 2) \left\{ \mathcal{H}_{\text{eff}}[\phi_\tau] \right\}_m]}
\end{equation}
satisfies $\tilde{U}_{\text{f}}[\phi_\tau]-\tilde{U}_{\text{f}}[\phi_{0i}]=U_{\text{f}}[\phi_\tau]-U_{\text{f}}[\phi_{0i}]$ hence is a useful candidate for a ``standalone'' fermionic potential. 
Another useful feature is $\tilde{U}_{\text{f}}[\phi_{0i}]=U_{\text{n}}[\phi_{0i}]$, 
thus the difference of subtracted fermionic and na\"ive potentials is simply $\tilde{U}_{\text{f}}[\phi_\tau]-U_{\text{n}}[\phi_\tau]$.
In what follows, one can thus envision Eq.~\eqref{eq:standalone} as a fermionic potential, and view the regularization in Eq.~\eqref{eq:ZPfaff} as providing a constant energy offset via $\tilde{U}_{\text{n}}[\phi_{0i}]$. 
In what follows, we refer to $\tilde{U}_{\text{f}}[\phi_\tau]$, together with this constant offset (to maintain consistency with previous sections) as our fermionic potential and drop the tilde to ease notations.

\subsection{Variational Instanton Strategy}
\label{subsec:strategy}

\begin{figure}
  \includegraphics[width=\linewidth]{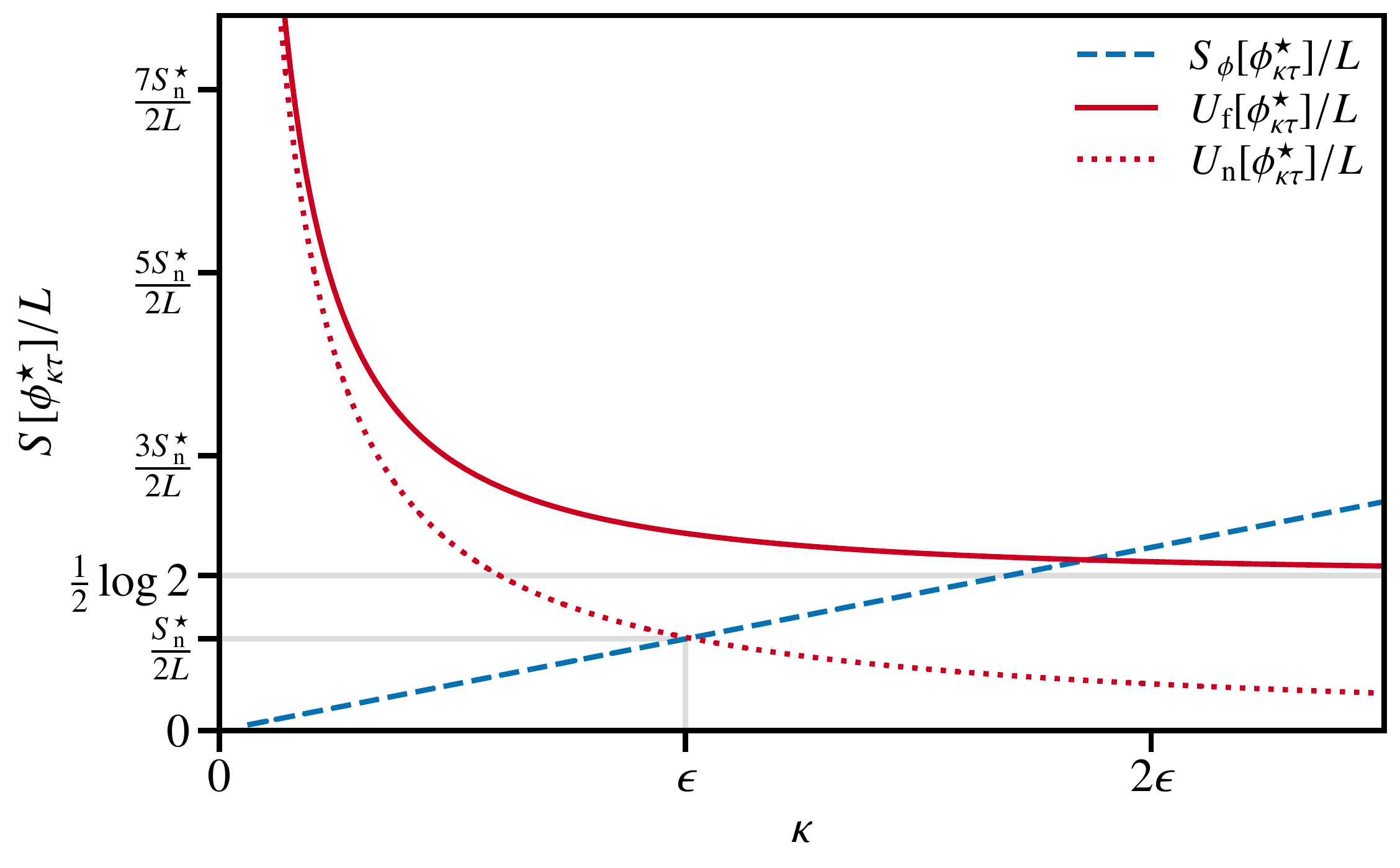}
  \caption{A typical plot of $U_\text{f}[\phi_{\kappa\tau}^{\star}] /L $ and $U_\text{n}[\phi_{\kappa\tau}^{\star}] / L$ against $\kappa$.
As $\kappa \to 0$, the fermionic potential tends to the na\"ive one, but as $\kappa \to \infty$ it approaches $\frac{1}{2} \log{2} + \mathcal{O}(1/L)$. 
Also plotted is the kinetic energy $S_{\phi} [\phi_{\kappa\tau}^{\star}]/L$.
  In terms of $\kappa$ and $\epsilon=\sqrt{E_\text{C}E_\text{J}}/\Delta$ (set to $\epsilon=1$ above), the kinetic and na\"ive potential energies of an individual instanton are $S_{\phi}[\phi_{\kappa \tau}^{\star}] = S^{\star}_{\text{n}} \kappa /(2\epsilon)\propto \kappa \Delta/E_\text{C}$ and $U_{\text{n}}[\phi_{\kappa \tau}^{\star}] = S^{\star}_{\text{n}} \epsilon/ (2 \kappa)\propto E_{\text{J}}/\kappa\Delta$, respectively, where $S^{\star}_{\text{n}}$ is the classical action given in Eq.~\eqref{eq:classical_action}.
}
  \label{fig:Uf_vs_Kinetic}
\end{figure}

The effective bosonic theory has a potential $U_{\text{f}}[\phi_\tau]$ that is nontrivial to evaluate even for a given profile $\phi_\tau$, and the dependence of the profile itself on $U_{\text{f}}[\phi_\tau]$ leads to an even more complex problem. 
We now outline a strategy to tackle this problem variationally.  
The crux of this strategy is to again expand the partition function as an instanton gas, motivated by the link we made in Sec.~\ref{sec:fermionic_factor} between the fermionic Pfaffian factor and the ground state potential.
The shapes of the classical instantons making up the gas are still the result of competition between kinetic and potential energy but with the new effective potential $U_{\text{f}}[\phi_\tau]$ their features may change compared with the na\"ive case.
To facilitate relating to our sharp-instanton results we use the instanton width as a variational parameter.
Specifically, we will allow the timescale of instantons to be different to their na\"ive counterparts $\phi_{\tau}^{\star}$ by considering scaled instantons $\phi_{\kappa \tau}^{\star}$ where we replace $\tau \to \kappa \tau$.
Since na\"ive instantons for different values of $\omega_0\propto \sqrt{E_\text{C}E_\text{J}}$ are themselves related to each other by rescaling [in the action, $E_\text{C}E_\text{J}$ sets merely the overall scale of the potential relative to $(\partial_\tau \phi)^2$], we can choose to define $\phi_{\tau}^{\star}$ as the na\"ive instanton for $\sqrt{E_\text{C}E_\text{J}}=\Delta$. (As this is just a reference classical configuration, it need not obey the $\sqrt{E_\text{C}E_\text{J}}\ll\Delta$ restriction.)
By defining $\phi_{\tau}^{\star}$ in this way, and optimizing over $\phi_{\kappa \tau}^{\star}$, the na\"ive action is minimized for $\kappa=\sqrt{E_\text{C}E_\text{J}}/\Delta$, the rescaling required to get from the reference instanton to the one with  $\omega_0\propto\sqrt{E_\text{C}E_\text{J}}$. 
We shall, of course, be optimizing not the na\"ive action but the one with $U_{\text{f}}[\phi_\tau]$. 
By using $\kappa$ as a variational parameter, allowed to deviate from $\sqrt{E_\text{C}E_\text{J}}/\Delta$, we will better approximate the new classical action in the presence of fermions, without comprehensively probing the large space of all possible instanton shapes.
Calculating the quantity $U_{\text{f}}[\phi_{\kappa \tau}^{\star}]$ for a range of $\kappa$ and comparing the minimal value of $S[\phi_{\kappa \tau}^{\star}] = S_{\phi}[\phi_{\kappa \tau}^{\star}] + U_{\text{f}}[\phi_{\kappa \tau}^{\star}]$ (which includes the kinetic energy $S_{\phi}[\phi_{\kappa \tau}^{\star}]$) 
to the minimal value of $S_{\text{n}}[\phi_{\kappa\tau}^{\star}]$
will finally give the correction to the tunneling amplitude.
(We also set $N_g=0$ to make $S_{\phi}[\phi_{\kappa \tau}^{\star}]$ real, knowing that the complex winding term $\exp{[-i \pi N_g w]}$ is accounted for later.)

From Fig.~\ref{fig:Uf_vs_Kinetic}, one can observe the key features of $U_{\text{f}}[\phi_{\kappa \tau}^{\star}]$ and $U_{\text{n}}[\phi_{\kappa \tau}^{\star}]$ as a function of $\kappa$. 
Taking $\kappa \to \infty$ corresponds to the sharp instanton limit, where $U_{\text{n}}[\phi_{\kappa \tau}^{\star}] \to 0$ and $U_{\text{f}}[\phi_{\kappa \tau}^{\star}] > 0$ was evaluated in Sec.~\ref{subsec:low_energy_analogy}.
(We provide another analytical derivation for this limit, using a different method, in Sec.~\ref{subsubsec:scattering_obc}.)
As the arguments in Sec.~\ref{subsec:low_energy_analogy} suggested, the difference $U_{\text{f}}[\phi_{\kappa \tau}^{\star}]-U_{\text{n}}[\phi_{\kappa \tau}^{\star}]$ is largest in this limit. 
The opposite ``adiabatic'' limit of $\kappa \to 0$ is most easily understood from Eq.~\eqref{eq:Lagrangian_alternative}.
From
$X^{\dagger}_{\phi_{\kappa\tau}} \partial_\tau X_{\phi_{\kappa \tau}} = \kappa X^{\dagger}_{\phi_{\tau}} \partial_{\tau} X_{\phi_{\tau}}$, we see that for $\kappa \to 0$ the last term becomes vanishingly small 
compared with $\oplus_{m} \varepsilon_{\phi_\tau m}^+ \sigma_3$ as this remains gapped for APBC and OBC, even if the gap is exponentially small in system size for OBC.
(We need not worry about the gap closing for PBC because the $k=0$ eigenstate does not evolve with $\phi$.)
Therefore, for increasingly slow instantons the fermionic potential tends to the na\"ive potential: $U_{\text{f}}[\phi_{\kappa \tau}^{\star}] \to U_{\text{n}}[\phi_{\kappa \tau}^{\star}]$ as $\kappa \to 0$.

Fig.~\ref{fig:Uf_vs_Kinetic} also shows that the first order condition $dS[\phi_{\kappa\tau}^\star]/d\kappa=0$ for minimizing the action yields similar $\kappa^\star$ whether one uses the fermionic or the na\"ive potential. 
Hence, $\sqrt{E_\text{C}E_\text{J}}/\Delta$ remains a good proxy for 
$\kappa^\star$.
Furthermore, since the kinetic term is the same for the fermionic and the na\"ive case, once the optimal value $\kappa^\star$ is found, the fermionic suppression will approximately be given by $U_{\text{f}}[\phi_{\kappa^\star \tau}^{\star}]-U_{\text{n}}[\phi_{\kappa^\star \tau}^{\star}]$.
Since both $U_{\text{f}}[\phi_{\kappa^\star \tau}^{\star}]$ and $U_{\text{n}}[\phi_{\kappa^\star \tau}^{\star}]$ are $\propto L$, topology-changing fermions suppress tunneling exponentially in $L$.

\subsection{Evaluating the Time-Ordered Exponential with Scattering Matrices}
\label{subsec:scattering_matrices}

We now present a method to numerically calculate $U_{\text{f}}[\phi_{\kappa\tau}^{\star}]$ for intermediate values of $\kappa$ and any boundary condition.
Recall that to compute the fermionic determinant, we need to evaluate the eigenvalues $\lambda_{nm}$ via the time-ordered exponential [cf.\ Eq.~\eqref{eq:eigenvalues_of_time_ordered}]
\begin{equation}
 \mathbf{M} (\beta,0) \equiv \exp \left( -\beta \mathcal{H}_{\text{eff}} [\phi_\tau] \right) 
 =  \mathcal{T} \exp \left[ -\int_0^\beta d \tau \, \mathcal{H}_{\phi_\tau} \right] .
\end{equation}
The time-ordered exponential can be evaluated numerically by discretizing the integral into $N$ steps
\begin{equation}
 \mathbf{M} (\beta,0) = \lim_{N\to \infty} \mathbf{M}_N \mathbf{M}_{N-1} \dots \mathbf{M}_1
 \label{eq:discrete_time_ordered_exponential}
\end{equation}
with $\mathbf{M}_n = \exp [-(\beta/N) \mathcal{H}_{\phi_{\tau_n}} ] $ and $\tau_n = \beta (n - 1/2)/N$.
Since $\mathbf{M}(\beta, 0)$ has both exponentially large and small eigenvalues~\cite{beenakker_random-matrix_1997}, the matrix product~\eqref{eq:discrete_time_ordered_exponential} is numerically unstable.

While matrix product~\eqref{eq:discrete_time_ordered_exponential} itself does not rely on chiral symmetry, our system does have this symmetry. 
This allows us to interpret each $\mathbf{M}_n$ as a transfer matrix that satisfies flux-conservation via $\sigma_1 \mathbf{M}_n^\dagger \sigma_1 = \mathbf{M}_n^{-1}$, which is ensured by the chiral symmetry of $\mathcal{H}_\phi (k) = - \sigma_1 \mathcal{H}_\phi (k) \sigma_1$.
This allows us to transform the product of transfer matrices~\eqref{eq:discrete_time_ordered_exponential} into a composition of scattering matrices, whose contraction is numerically more stable~\cite{Tamura:1991ki}.

The reformulation of the time-ordered exponential as a scattering problem has the further advantage of simplifying the expressions we are ultimately interested in.
For profiles symmetric around $\beta/2$, i.e., $\phi_{\beta/2-\tau} = \phi_{\beta/2 + \tau}$, corresponding to an instanton-anti-instanton pair, the transfer matrices $\mathbf{M}(0,\beta/2)$ and $\mathbf{M}(\beta/2,\beta)$ are related via imaginary time reversal, $\mathbf{M}(\beta/2,0) =[\mathbf{M}(\beta,\beta/2)]^\dagger$, which relates their respective scattering matrices $\mathbf{S}(\beta/2,0) = - \sigma_2 [\mathbf{S}(\beta,\beta/2)]^\dagger \sigma_2$.
A straightforward calculation using the polar decomposition reveals that the transmission eigenvalues of the full scattering matrix are $T_{\mathrm{full},m} = T_m^2/(2-T_m)^2$, where $T_m$ are the transmission eigenvalues of $\mathbf{S}(\beta/2,0)$.
The transmission eigenvalues are related to $e^{\pm x_m}$, the eigenvalues of $\mathbf{M}(\beta,0)$ with real $x_m$, via $T_{\mathrm{full},m} = 1/\cosh^2 x_m$~\cite{beenakker_random-matrix_1997}.
Since $\pm x_m/\beta$ are the eigenvalues of the effective Hamiltonian $\mathcal{H}_{\text{eff}} [\phi_\tau]$, the fermionic determinant is thus proportional to the product of all [cf.\ Eq.~\eqref{eq:determinant_product}]
\begin{equation}
 \cosh \left(\frac{1}{2} x_m  \right) =  \cosh \left[ \frac{1}{2} \arccosh \left( \frac{2-T_m}{T_m} \right) \right] = \frac{1}{\sqrt{T_m}} ,
 \label{eq:product_of_coshs}
\end{equation}
i.e., the fermionic determinant for such symmetric configurations is proportional to $1/\det (\mathbf{t})$, where $\mathbf{t}$ is the transmission matrix for \emph{half} of the imaginary time evolution, consisting of \emph{one} instanton.

\subsubsection{Analytic Results for Sharp Instantons with OBC}
\label{subsubsec:scattering_obc}

The scattering matrix formalism also allows us to compute the fermionic determinant analytically in the sharp-instanton limit, including for OBC, without explicitly referring to the chiral boundary modes.
We first rotate the fermionic Hamiltonian~\eqref{eq:fermionic_hamiltonian} $\mathcal{H}_\phi \to \mathcal{H}_\phi'$ via $\sigma_3 \to \sigma_1$, giving
\begin{align}
 \mathcal{H}_\phi' = \begin{pmatrix} & A_{\phi}^{} \\ A_{\phi}^\dagger & \end{pmatrix}, & & 
 A_{\phi}^{}(k) = \Delta [\cos^2 (\phi/2) - \sin^2 (\phi/2) e^{-ik} ] .
\end{align}
Using the singular value decomposition $A_{\phi} = W_\phi \Sigma_\phi Y_\phi^\dagger$, each transfer matrix for a $\delta\tau$ slice can be brought into its polar form~\cite{Mello:1988cj,Martin:1992ea}, hence each scattering matrix is
\begin{equation}
 \mathbf{S} = \begin{pmatrix} - Y_{\phi}^{} & \\ & W_{\phi}^{} \end{pmatrix} \begin{pmatrix} -\tanh (\delta\tau \Sigma_{\phi}) & \sech(\delta\tau \Sigma_{\phi}) \\ \sech (\delta\tau  \Sigma_{\phi}) & \tanh (\delta\tau  \Sigma_{\phi}) \end{pmatrix} \begin{pmatrix} W_{\phi}^\dagger & \\ & -Y_{\phi}^\dagger \end{pmatrix}.
 \label{eq:chiral_scattering_matrix}
\end{equation}
For a system of size $L$ with OBC each sub-block is an $L \times L$ matrix.
At $\phi =0$, we consider the modified chemical potential $\Delta \to \Delta'=\Delta (1-1/L)$ to ensure that the ground state energies match (cf.\ Sec.~\ref{subsubsec:low_energy_analogy_obc}).
The singular value decomposition at $\phi=0$ is trivial ($\Sigma_0 = \Delta'$ with $W_0 = Y_0 = 1$), and at $\phi=\pi$ yields $W_\pi = -1$, $\Sigma_\pi = \Delta \diag (0,1,\dots 1)$ and
\begin{equation}
 Y_\pi^\dagger = \begin{pmatrix}
  0 & & & 1 \\
  1 & 0 & & \\
  & \ddots & \ddots & \\
  & & 1 & 0
 \end{pmatrix} .
\end{equation}
The sharp instanton limit of a symmetric instanton-anti-instanton configuration, where according to Eq.~\eqref{eq:product_of_coshs} one may consider just the instanton, has two scattering matrices: one for imaginary time interval $[0,\beta/4)$ at $\phi=0$ and another $[\beta/4,\beta/2)$ at $\phi=\pi$. 
Their contraction~\cite{Tamura:1991ki} gives the lower triangular transmission matrix
\begin{align}
 \mathbf{t} = \frac{1}{\cosh \left( \frac{\beta\Delta'}{4} \right) \cosh \left( \frac{\beta\Delta}{4} \right)} \begin{pmatrix}
  \cosh \left( \beta\Delta / 4 \right) & & & & \\
  y & 1 & & & \\
  y^2 & y & 1 & & \\
      & \ddots & \ddots & \ddots & \\
  y^{L-1} & \dots & y^2 & y & 1
 \end{pmatrix},
\end{align}
with $y = \tanh (\beta \Delta/4) \tanh (\beta \Delta'/4)$.
Its product of singular values
\begin{align}
 \prod_m & \sqrt{T_m} = | \det \mathbf{t}_\mathrm{tot} | = \frac{\cosh \left( \beta\Delta / 4 \right)}{\left[ \cosh \left( \beta\Delta'/4 \right) \cosh \left( \beta\Delta / 4 \right) \right]^{L}}
\end{align}
equals the determinant of the transmission matrix.

The product of singular values is proportional to the square root of the fermionic determinant [Eqs.~\eqref{eq:determinant_product} and~\eqref{eq:product_of_coshs}], which gives for the ratio of a two-instanton and zero-instanton configuration in the sharp-instanton limit
\begin{align}
 \sqrt{ \frac{\det{[\partial_\tau + \mathcal{H}_{2\text{i}}]}}{\det{[\partial_\tau + \mathcal{H}_{0\text{i}}]}} } = \frac{ \left[ \cosh \left( \beta \Delta' / 2 \right) \right]^L}{\left[ \cosh \left( \beta\Delta' / 4 \right) \right]^L \left[ \cosh \left( \beta \Delta / 4 \right) \right]^{L-1}},
\end{align}
and, for $\beta \Delta \gg 1$,
\begin{equation}\label{eq:inst_scatt}
 \sqrt{ \frac{\det{[\partial_\tau + \mathcal{H}_{2\text{i}}]}}{\det{[\partial_\tau + \mathcal{H}_{0\text{i}}]}} } = 2^{L-1} \left[ 1 + O \left( e^{-\left( 1-\frac{1}{L}\right) \frac{\beta\Delta}{2}} \right) \right].
\end{equation}
Therefore, when the instanton separation $\beta/2$ is well beyond the width  $\Delta^{-1}$ of the instanton-bound fermionic edge mode in the temporal direction ($\Delta^{-1}$ is the ``temporal coherence length''  owing to the temporal velocity equaling unity), the determinant ratio does not depend on the instanton separation. 
Note that, as in Sec.~\ref{subsubsec:low_energy_analogy_obc}, this OBC result relies on the matching ground state energies at $\phi=0,\pi$; for values of $\Delta'$ other than $\Delta (1 - 1/L)$, the ratio generally grows exponentially with instanton separation. 
Eq.~\eqref{eq:inst_scatt} agrees with the result Eq.~\eqref{eq:suppression_edge_modes} from the boundary-mode approach.

\subsection{Results}
\label{sec:results}

\begin{figure}
  \includegraphics[width=\linewidth]{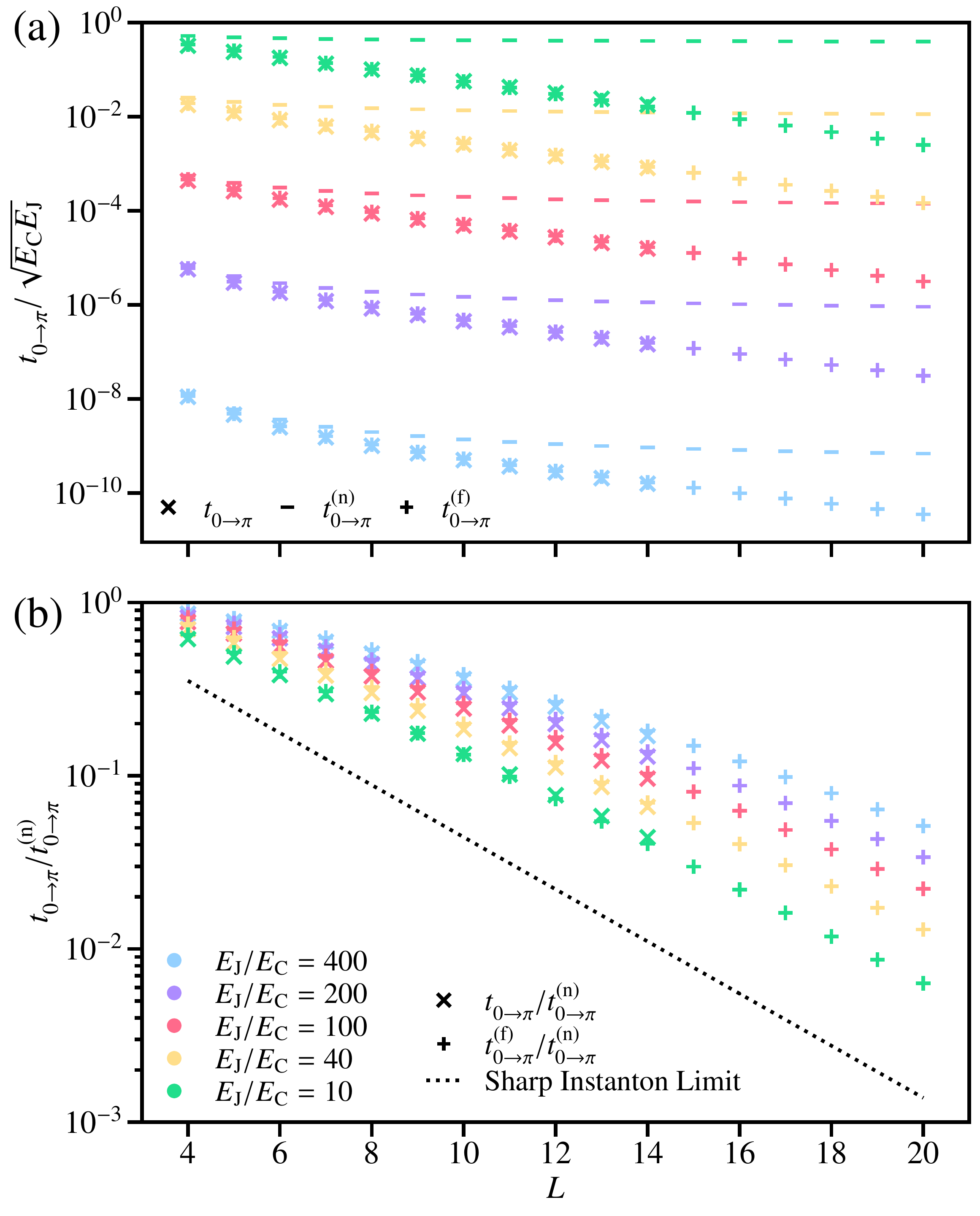}
  \caption{Scaling of the tunneling amplitude for OBC systems of size $L$, plotted on a logarithmic scale for different ratios of $E_{\text{J}} / E_{\text{C}}$.
  The effective Josephson energy is defined as $E_\text{J} = L\Delta (\pi - 2)/4\pi$ for simplicity.
  (a) The tunneling amplitude predicted by our instanton-based approach [$t_{0\to\pi}^{(\text{f})}$] and by the energy splitting obtained from exact diagonalization [$t_{0\to\pi}^{}$], in contrast to the na\"ive approach [$t_{0\to\pi}^{(\text{n})}$] that takes only the fermionic ground state energy into account.
  Although plotted up to $L=20$, the instanton calculation can easily estimate the tunneling suppression for even larger systems, whereas exact diagonalization quickly becomes impractical.
  (b) The ratio of exact and na\"ive results, compared with our prediction.
  We also plot the sharp instanton limit results (dotted line), which bound the suppression of the tunneling amplitude.}
  \label{fig:OBC_scaling}
\end{figure}

We now calculate the tunneling suppression due to the topology-changing fermions obtained from our variational approach. 
To test our results, we will compare with the splitting of the lowest two energies obtained by exact diagonalization.
We focus on OBC. (Results for APBC are given in Appendix~\ref{sec:other_BCs}.)

As we noted in Secs.~\ref{sec:model} and \ref{sec:instantons}, $E_\text{J}/E_\text{C}$ and $\sqrt{E_\text{J}E_\text{C}}/\Delta \propto \omega_0/\Delta$ are two key dimensionless parameters of the problem. 
The small tunneling, i.e., semiclassical, limit is $E_\text{J}/E_\text{C}\gg 1$; this is also the criterion for instanton methods to be valid (cf.\ Sec.~\ref{subsec:instanton_gas}).
Conversely, working in the regime where the fermionic gap near the potential minima is operative (in the sense of $\phi$'s dynamics) requires $\sqrt{E_\text{J}E_\text{C}}/\Delta \ll 1$. 

Since the scale of the Josephson potential $E_{\text{J}} \approx (V_{\pi/2} - V_{0}) / 2$ grows linearly with $L$, we require different scaling of $E_\text{C}$ depending on which of the key dimensionless parameters we keep fixed in our calculations:  
one needs $E_\text{C}\sim L^{-1}$ (a scaling that can naturally arise in planar Josephson junction systems) to keep $\omega_0/\Delta$ fixed, while $E_\text{C}\sim L$ is required for fixing $E_\text{J}/E_\text{C}$. 

To assess the performance of our variational calculation, below we focus on fixing the parameter $E_\text{J}/E_\text{C}$ characterizing the adequacy of the semiclassical limit. 
In using this parameter, we must bear in mind however, that now $\sqrt{E_\text{J}E_\text{C}}\sim L$ hence we must keep $L$ finite to remain in the $\omega_0 \ll \Delta$ regime.

In our numerical exact diagonalization of the full Hamiltonian~\eqref{eq:charging_BdG}, we work in the charge basis.
Terms proportional to $\cos{\phi}$ are off-diagonal in this basis, $e^{i\phi} = \sum_N \ket{N+2} \bra{N}$.
Owing to the  $\sim(N - N_g)^2$ charging term, only a certain number of charge states centered around $N_g$ contribute to the ground state, hence the charge basis can be truncated to $N_s$ states and the low-energy spectrum will still converge to acceptable accuracy.
The na\"ive bosonic problem, where we replace the fermionic Hamiltonian by its ground state energy, is also solved by diagonalizing a Hamiltonian with a truncated basis.

As suggested by the form of the low-energy dispersion $E_{\pm}(N_g)$ in Eq.~\eqref{eq:tight_binding_spectrum}, the desired tunneling amplitude is observable from the energy splitting when the wells are symmetric [diagnosed by the condition $E_+(1) = E_-(1)$] and is given by
\begin{equation}
  t_{0\to\pi} = \left[ E_+(0) - E_-(0) \right] \big/ 2.
\end{equation}
However, even after the symmetrization procedure in Secs.~\ref{subsubsec:low_energy_analogy_obc} and~\ref{subsubsec:scattering_obc}, the curvature of the two wells is different.
To counter this effect, we add another Josephson potential to ensure that the harmonic-oscillator-like states in both wells would have the same energy were it not for tunneling.

In Fig.~\ref{fig:OBC_scaling}(a), we show the tunneling amplitude $t_{0\to\pi}^{\text{(f)}}$ based on the instanton calculation and, for small systems up to $L=14$, the tunneling amplitude $t_{0\to\pi}$ from the energy splitting that we computed by exact diagonalization.
We compare these results with the na\"ive tunneling $t_{0\to\pi}^{\text{(n)}}$. 
While $t_{0\to\pi}^{\text{(n)}}/\sqrt{E_\text{J}E_\text{C}}$ quickly approaches an $L$-independent value, the tunneling amplitude $t_{0\to\pi}$ decreases exponentially with $L$.
The instanton-based result $t_{0\to\pi}^{\text{(f)}}$ and the exact $t_{0\to\pi}$ almost coincide.

To highlight the suppression by the fermionic contribution, we compare the ratios $t_{0\to\pi}^{\text{(f)}} / t_{0\to\pi}^{\text{(n)}}$ and $t_{0\to\pi}^{} / t_{0\to\pi}^{\text{(n)}}$ in Fig.~\ref{fig:OBC_scaling}(b).
The suppression gets weaker with larger $E_{\text{J}} / E_{\text{C}}$ and would eventually approach the na\"ive result.
This can be understood by noting that for fixed $E_{\text{J}} / E_{\text{C}}$,  and due to $E_\text{J} \propto L\Delta$ in our system, we have $\sqrt{E_{\text{J}}E_{\text{C}}}/\Delta \propto L \sqrt{E_{\text{C}}/E_{\text{J}}}$. 
Therefore, larger $E_{\text{J}} / E_{\text{C}}$ leads to smaller $\omega_0$ which implies larger instanton width, and hence smaller optimal value $\kappa^\star$. 
The smaller $\kappa^\star$ the more the fermionic and  na\"ive potentials are alike (Fig.~\ref{fig:Uf_vs_Kinetic}), and, since the fermionic suppression is due to the difference between these two potentials, the closer we are to the na\"ive result. 
Conversely, for small $E_{\text{J}} / E_{\text{C}}$, the tunneling suppression approaches the upper bound derived from the sharp instanton limit.
Upon increasing $L$, the $\omega_0\propto L$ dependence, by narrowing instantons and hence increasing $\kappa^\star$,  also pushes $t_{0\to\pi}^{\text{(f)}} / t_{0\to\pi}^{\text{(n)}}$ towards the sharp instanton limit; this leads to a slight downward bend in $t_{0\to\pi}^{\text{(f)}} / t_{0\to\pi}^{\text{(n)}}$ as a function of $L$. 

While a fuller estimate would require evaluating the fermionic Pfaffian factor beyond classical instanton configurations, we see that using just the classical configuration works remarkably well.
One would anticipate more deviation from our prediction in parameter regimes departing from the semiclassical regime $E_{\text{J}} / E_{\text{C}}\gg1$, where fluctuations in the path integral give a greater contribution to the tunneling amplitude~\cite{altland_condensed_2010}.
For numerically accessible system sizes, an exponential fit to instanton and exact diagonalization results produces the same fermionic suppression scaling exponent   (within the standard error of the fit) for each $E_\text{J} / E_\text{C}$ series, with only a small offset.
This agreement persists across a wide parameter range, but it becomes worse with smaller $E_{\text{J}} / E_{\text{C}}$ ratios as is expected upon gradually departing from the semiclassical regime. 
While, to maintain $\sqrt{E_{\text{J}}E_{\text{C}}} \lesssim \Delta$ (with $\sqrt{E_{\text{J}}E_{\text{C}}}\approx0.57\Delta$ for $E_{\text{J}} / E_{\text{C}}=10$ and $L=20$), Fig.~\ref{fig:OBC_scaling} focuses on moderate $L$, the range considered already  emphasizes that the instanton calculation allows for the treatment of system sizes well beyond the reach of exact diagonalization.

\section{Conclusion}
\label{sec:conclusions}

In this work, we studied how coupling to a fermionic bath impacts the tunnel amplitude of a particle, if the tunneling between potential minima, where the bath is gapped, requires a change in fermionic topology and hence a gap closing. 
In general, for fermions in $d$ dimensions, we used the field theoretical language of instantons to map this tunneling problem to that of interfaces between topologically distinct regions in $d+1$ dimensions. 
This relation, as we elucidated in Sec.~\ref{subsec:low_energy_analogy}, amounts to stepping up on a dimensional ladder, akin to the reversal of topological insulators' and superconductors' dimensional reduction procedures discussed in Refs.~\cite{kitaev2009periodic,Ryu_2010,QiZhang_RevModPhys.83.1057,bernevig2013topological}. 
The existence of topologically protected gapless boundary modes in these $(d+1)$-dimensional geometries leads to a suppression of tunneling amplitude compared to the value one would na\"ively expect by taking the bath at its instantaneous ground state. This suppression is exponential in the size of the fermionic system. 
We demonstrated this in detail on our $d=1$ example, including establishing an analytical bound setting out the strongest possible fermionic suppression. 
This bound corresponds to sharp instantons, a tractable scenario also applicable to $d>1$ where it is expected to lead to analogous results: an exponential suppression with $L^d$,  with the exponent set by the boundary modes' density of states. 

Complementary to this picture, we also showed how to use instanton field theory to incorporate topology-changing fermions into a variational calculation. 
This method, which also revealed an unexpected link to scattering matrices that usually arise in quantum transport calculations, allowed us to probe a range between wide instantons (no fermionic suppression) and sharp instantons (maximal fermionic suppression). 
We compared the tunneling amplitude obtained from this variational path-integral method with the energy splitting computed by exact diagonalization of the full many-body system.
Our method uses only one variational parameter (the instanton width), and this already yields results that match excellently with exact diagonalization, while being able to reach much larger systems sizes. 
In particular, while we demonstrated its use on our $d=1$ system, the method is equally well applicable to higher dimensions where exact diagonalization would be limited to exceedingly small systems. 

Although we focused on conceptual aspects, our results may be relevant for the planar Josephson systems~\cite{pientka_topological_2017,Hell:2017fn,Fornieri:2019hz,Ren:2019ew} that served as inspiration. 
As in our $d=1$ model, the key dimensionless parameters are $E_\text{J}/E_\text{C}$ and $\omega_0/\Delta \sim \sqrt{E_\text{J}E_\text{C}}/\Delta$ with $\Delta$ the induced superconducting gap. 
(Large $E_\text{J}/E_\text{C}$ again corresponds to the semiclassical regime where instanton methods are expected to work, while $\sqrt{E_\text{J}E_\text{C}}\ll \Delta$ renders the fermionic gap operative near the potential minima.)
In these systems, the effective Josephson energy $\propto L$ and the charging energy  $\propto 1/L$ (being inversely proportional to capacitance). 
Hence, $\omega_0/\Delta$ is fixed thus, unlike the fixed $E_\text{J}/E_\text{C}$ case we used for assessing our variational method,  the large $L$ regime can be taken consistently with $\omega_0\ll\Delta$.  
Although due to $\sqrt{E_\text{J}/E_\text{C}} \propto L$ even the na\"ive tunneling amplitude is suppressed exponentially, we stress that the suppression we found enhances the tunneling exponent.
(In other setups, it may be possible to have kinetic and na\"ive potential terms that do not scale with the size $L$ of the fermionic bath; then one may have an $L$-independent na\"ive tunneling exponent, together with fixed $\omega_0$ and thus a consistent large $L$ limit, and an exponential-in-$L$ suppression solely from fermionic topological effects.)

Since the fermionic ground state energies in trivial and nontrivial regimes are not necessarily equal, the observation of the fermionic suppression of the tunneling amplitude via the energy splitting  may be challenging in these Josephson systems.
However, the tunneling amplitude also impacts non-equilibrium effects which may be more amenable for observation in experiments. 
In investigating these and other features, studying local versions of our model (obtained by incorporating $\partial_x\phi$) may offer a useful direction for the future.  

The fermionic tunneling suppression we found may be relevant for considering combining topological and transmon qubits, as for example when applying schemes that utilize the charging energy for braiding and parity readout~\cite{Jiang:2011cs,Bonderson:2011bc,Hassler:2011gj,vanHeck:2012bp,Hyart:2013gf} to planar Josephson junctions~\cite{pientka_topological_2017,Hell:2017fn}.
The tunneling suppression could also potentially be used to better suppress phase slips (and thus charge noise) in transmon qubits~\cite{Koch:2007gz,Schreier:2008gs}.

The fact that tunneling is only suppressed (but not completely blocked) between topologically distinct minima is also suggestive of the prospects to realize quantum superpositions between topologically distinct fermionic ground states.
This is especially intriguing for OBC, where, as in our $d=1$ model, it can translate to superpositions of fermionic many-body states with and without Majorana end modes.
Owing to the exponentially localized nature of these Majorana end states, and to their localization exponent being unrelated to that of the tunnel suppression, these end states can meaningfully exist in moderate-sized systems where tunneling between topologically distinct minima can play a considerable role. 

\begin{acknowledgments}
This work was supported by an EPSRC Studentship, the ERC Starting Grant No. 678795 TopInSy and the EPSRC grant EP/S019324/1.
\end{acknowledgments}

\appendix

\section{Jackiw-Rebbi Derivation of Edge Mode Spectrum}
\label{sec:Jackiw-Rebbi}

In this Appendix, we derive the edge mode spectrum associated with an instanton using a Jackiw-Rebbi-like ansatz~\cite{Jackiw:1976ky}.
We shall use the momentum-space representation $\mathcal{H}_{\phi}(k)$ of the 1D model [Eq.~\eqref{eq:fermionic_hamiltonian}] to write the Lagrangian 
\begin{equation}
  \mathcal{L}(\tau, k) = \partial_{\tau} + \mathcal{H}_{\phi_{\tau}}(k).
\end{equation}
Recall that a sign change of $\cos{\phi_\tau}$ corresponds to a topological phase transition of $\mathcal{H}_{\phi_{\tau}}(k)$.
As in the main text, we consider the Hermitian Hamiltonian $\tilde{\mathcal{H}}(\tau, k)=i\sigma_{1}\mathcal{L}(\tau, k)$.

We deal solely with the case of unequal gaps on both sides of the transition because this encompasses the case of equal gaps.
To this end, we modify the gap at $\phi=0$ as $\Delta \to \Delta' = \eta \Delta$, where $\eta \in (0, 1] $ is a parameter describing the asymmetry of the gap.
Separating out the $\phi_\tau$ dependence, the Hermitian Hamiltonian is now
\begin{align}
  \tilde{\mathcal{H}}(\tau, k) =& \frac{\Delta}{2}\cos{\phi_{\tau}} \left[\left(\eta+\cos{k}\right)\sigma_{2}-\sin{k}\sigma_{3}\right] \nonumber \\
  &+\frac{\Delta}{2}\left[\left(\eta-\cos{k}\right)\sigma_{2}+\sin{k}\sigma_{3}\right]+i\partial_{\tau}\sigma_{1}.
  \label{eq:2D_Hamiltonian_asymmetric}
\end{align}
Suppose that there is an instanton located at $\tau_0$ which closes the gap: $\cos{\phi_{\tau < \tau_0}} > 0$ and $\cos{\phi_{\tau > \tau_0}} < 0$.
One might propose an ansatz
\begin{equation}
  \Psi\left(k,\tau\right) \stackrel{?}{=} \exp\left[\Delta \cos{(k / 2)}\int_{\tau_{0}}^{\tau}d\tau^{\prime}\cos\phi_{\tau^{\prime}}\right]\psi_k
  \label{eq:Jackiw-Rebbi_ansatz}
\end{equation}
localized at $\tau_0$, where the sign change of $\cos{\phi_\tau}$ ensures that the solution remains normalizable on both sides of the transition.
Such an ansatz fails for $\eta \neq 1$ because the decay of the bound state needs to be different in regions with a different gap~\cite{Charmchi:2014eu,Jana:2019eb}.
We therefore try a judicious rewriting of the Hamiltonian~\eqref{eq:2D_Hamiltonian_asymmetric} that immediately suggests a better ansatz, namely
\begin{align}
  \tilde{\mathcal{H}} =&i\partial_{\tau}\sigma_{1} +  \frac{\Delta}{2} ( \alpha_k + \cos{\phi_{\tau}}) \left[\left(\eta+\cos{k}\right)\sigma_{2}-\sin{k}\sigma_{3}\right]  \nonumber \\
  + & \frac{\Delta}{2}\left[\left(\eta(1-\alpha_k)-(1+\alpha_k)\cos{k}\right)\sigma_{2}+(1+\alpha_k)\sin{k}\sigma_{3}\right],
\end{align}
where we introduce a parameter $\alpha_k$ to label the reshuffling.
We will soon see that only one choice of $\alpha_k$ makes the ansatz work.
This new form suggests the ansatz
\begin{equation}
  \Psi =\exp\left[\frac{\Delta}{2} \sqrt{1+\eta^2+2\eta\cos{k}}\int_{\tau_{0}}^{\tau}d\tau^{\prime}(\alpha_k + \cos{\phi_{\tau^{\prime}})}\right]\psi_k
  \label{eq:Jackiw-Rebbi_ansatz_asymmetric}
\end{equation}
that factorizes the Hamiltonian as
\begin{equation}
  \tilde{\mathcal{H}} \psi_k = \left[i\Delta (\alpha_k + \cos{\phi_{\tau}}) \sqrt{1+\eta^2+2\eta\cos{k}} \sigma_1 Q_k + h_k \right] \psi_k.
\end{equation}
with the projector
\begin{equation}
  Q_k = \frac{1}{2} \left[\sigma_0 + \frac{\left(\eta+\cos{k}\right)\sigma_{3}+\sin{k}\sigma_{2}}{\sqrt{1+\eta^2+2\eta\cos{k}}} \right] = Q_k^2,
\end{equation}
and the $\tau$-independent term
\begin{equation}
  h_k = \frac{\Delta}{2}\left[ \left(\eta(1-\alpha_k) \mkern1.5mu{-}\mkern1.5mu (1+\alpha_k)\cos{k}\right)\sigma_{2} \mkern2.5mu{+}\mkern2.5mu (1+\alpha_k)\sin{k}\sigma_{3} \right].
\end{equation}

\begin{figure}[h!]
  \includegraphics[width=\linewidth]{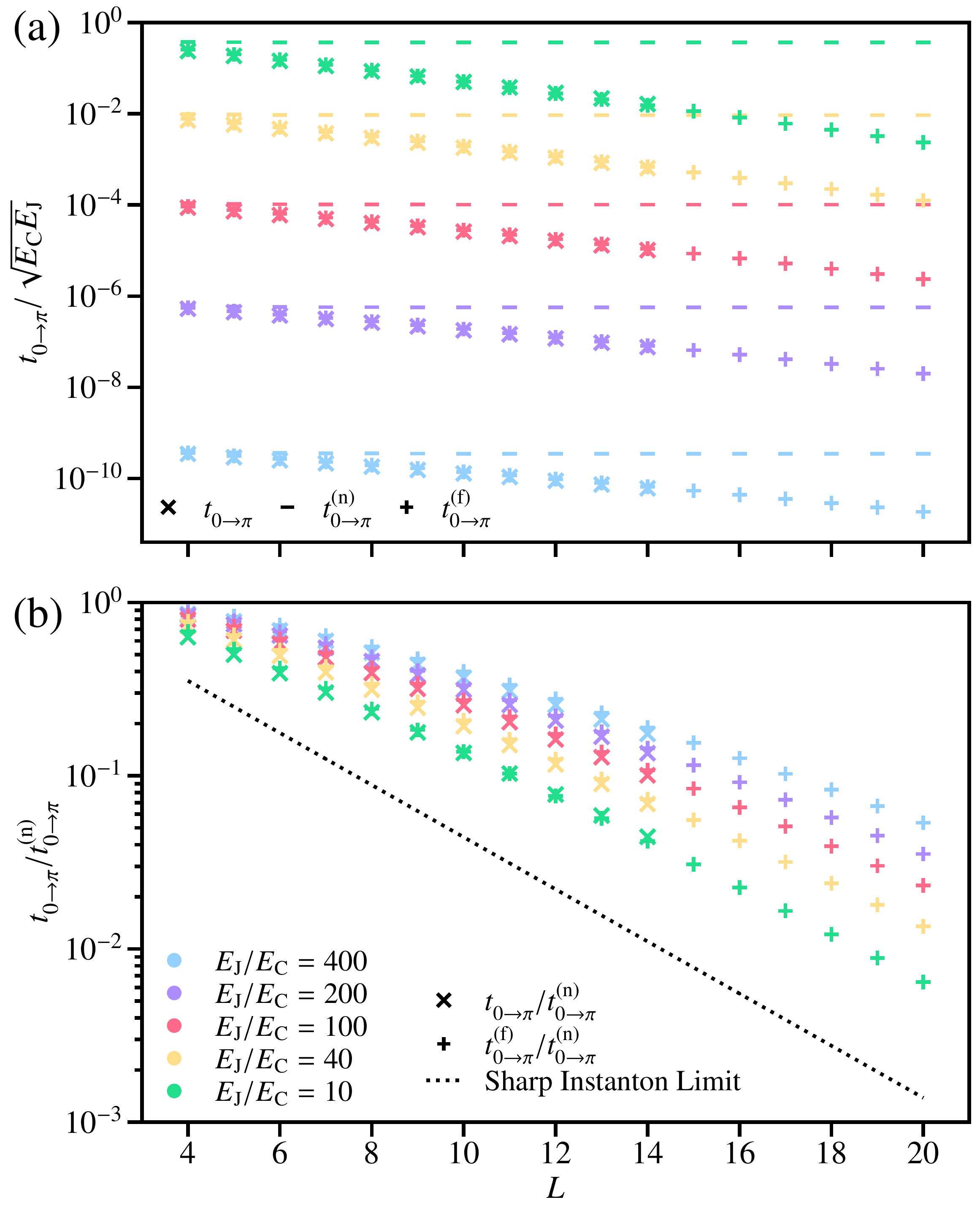}
  \caption{Scaling of the tunneling matrix element $t_{0\to\pi}$ for APBC systems of size $L$, plotted for different ratios of $E_{\text{J}} / E_{\text{C}}$ as we did for OBC in Fig.~\ref{fig:OBC_scaling}.
  (a) The tunneling amplitudes $t_{0\to\pi}$ (divided by the plasma frequency $\sqrt{E_{\text{J}} E_{\text{C}}}$) calculated with different approaches.
  (b) The ratio between the exact and na\"ive results in the plot above, compared with that predicted by the instanton calculation.}
  \label{fig:APBC_scaling}
\end{figure}

To get $\tau$-independent solutions we project onto the $Q_k \psi_k^{+} = 0$ subspace in which the Hamiltonian is simply $\tilde{\mathcal{H}}\psi_{k}^{+} = h_k \psi_{k}^{+}$.
For this to be valid, we need $\psi_{k}^{+}$ to be a simultaneous eigenstate of both the projector and the remaining effective Hamiltonian, i.e. $[Q_k, h_k] = 0$, which holds if
\begin{equation}
  \alpha_k = \frac{\eta^2 - 1}{1+\eta^2+2\eta\cos{k}}.
\end{equation}
A crucial observation is that $\alpha_k$ monotonically decreases from $\alpha_{k=0} = (\eta - 1)/(\eta + 1)$ to $\alpha_{k=\pi} = (\eta + 1)/(\eta - 1)$, which means that there exists a range of $k$ for which $\alpha_k < -1$ and the ansatz of Eq.~\eqref{eq:Jackiw-Rebbi_ansatz_asymmetric} is no longer normalizable.
Thus bound states only exist in the range $|k| \leq \arccos{(-\eta)}$, for which the dispersion is given by
\begin{equation}
  E_{\parallel}^{\pm}(k) = \frac{\Delta}{2} \sqrt{(1+\alpha_{k})^2 + \eta^2 ( 1-\alpha_{k})^2 - 2\eta( 1 - \alpha^2_{k})\cos k},
\end{equation}
saturating at the value of the reduced gap $\Delta' = \eta \Delta$.
Setting $\eta = 1$ recovers the equal gap case, which has bound states for all $k$ with simple dispersion
\begin{equation}
  E_{\parallel}^{\pm}(k) = \Delta \sin{(k/2)}
\end{equation}
quoted in the main text.
We have thus derived the spectrum of the chiral edge mode along the spatial direction, bound to each instanton.
Had $\cos{\phi_\tau}$ changed sign in the opposite direction (as for an anti-instanton), the ansatz in Eq.~\eqref{eq:Jackiw-Rebbi_ansatz_asymmetric} would need a minus sign in the exponent to be normalizable, and we would have derived an edge mode of opposite chirality.

Recall from Sec.~\ref{subsubsec:low_energy_analogy_obc} that for the OBC case to have symmetric wells, one tunes the gap inequality parameter to be $\eta=1-\frac{1}{L}$, which we may substitute into the above expressions to find the chiral edge mode spectrum associated with each instanton.

\section{Numerical Results for APBC}
\label{sec:other_BCs}

We present numerical results for APBC in Fig.~\ref{fig:APBC_scaling}, which is the analogue to Fig.~\ref{fig:OBC_scaling} from the main text.
We see that the modified instanton calculation performs equally well for APBC as for OBC.
The results for APBC are qualitatively similar to OBC, consistently with the expectation based on our sharp-instanton considerations (Sec.~\ref{subsubsec:low_energy_analogy_obc}).

\bibliography{majorana-jj}

\end{document}